\documentclass[letterpaper,english,aps,prb,twocolumn,floatfix, showpacs, amsfonts, amssymb, superscriptaddress]{revtex4}
\usepackage[T1]{fontenc}
\usepackage[latin1]{inputenc}
\usepackage{amsmath}
\usepackage{babel}
\usepackage{graphics}
\usepackage{amssymb}
\usepackage{subfigure}

\makeatletter

\usepackage{amscd}
\usepackage{bm}

\makeatother
\begin{document}

\title{A Coulomb gas approach to the anisotropic one-dimensional Kondo lattice
model at arbitrary filling}

\author{E.~Novais}

\email{enovais@bu.edu}

\affiliation{Instituto de Física Gleb Wataghin, Unicamp, Caixa Postal 6165, 13083-970
Campinas, SP, Brazil}
\affiliation{Department of Physics, Boston University, Boston, MA 02215}

\author{E.~Miranda}

\email{emiranda@ifi.unicamp.br}

\affiliation{Instituto de Física Gleb Wataghin, Unicamp, Caixa Postal 6165, 13083-970
Campinas, SP, Brazil}

\author{A.~H.~Castro~Neto}

\email{neto@buphy.bu.edu}

\affiliation{Department of Physics, Boston University, Boston, MA 02215}

\author{G.~G.~Cabrera}

\email{cabrera@ifi.unicamp.br}

\affiliation{Instituto de Física Gleb Wataghin, Unicamp, Caixa Postal 6165, 13083-970
Campinas, SP, Brazil}

\date{\today}

\begin{abstract}
We establish a mapping of a general spin-fermion system in one dimension
into a classical generalized Coulomb gas. This mapping allows a renormalization
group treatment of the anisotropic Kondo chain both at and away from
half-filling. We find that the phase diagram contains regions of paramagnetism,
partial and full ferromagnetic order. We also use the method to analyze
the phases of the Ising-Kondo chain.
\end{abstract}


\maketitle

\section{Introduction}

The relevance of studying the Kondo lattice model (KLM) has not
decreased.  It is believed to be at the heart of the physics of both
the heavy fermion materials, in its antiferromagnetic (AFM)
version,\cite{Hewson} and also the manganites, when local moments and
conduction electrons interact via the ferromagnetic (FM) Hund's
coupling.\cite{Dagotto1} In the first case, the well understood
behavior of the single (or few) impurity Kondo model permeates much of
our current understanding.  However, the interplay between the local
Kondo physics and the non-local RKKY interaction in a lattice
environment\cite{Doniach} remains elusive in current approximate
schemes, although it may play a prominent role close to quantum
critical points\cite{JAH76,AJM93,Coleman,QSSR+01} or even away from
them. In this respect, a more thorough understanding of the
one-dimensional (1D) case might be fruitful, even in light of the
peculiarities of 1D systems. Furthermore, the study of the 1D KLM is
important in its own right for the analysis of some
quasi-one-dimensional organic compounds such as $\rm (Per)_2M(mnt)_2$
(M=Pt,Pd), \cite{bourbonnais,lopes,matos} and $\rm
(DMET)_2FeBr_4$.\cite{enoki}

A fairly complete ground state phase diagram has been established for
the 1D KLM.\cite{Tsunetsugu, Dagotto1} On the one hand, the
antiferromagnetic model at half-filling has both charge and spin
gaps.\cite{Tsvelik1} For lower band fillings there is a quantum phase
transition from a paramagnetic ground state to a ferromagnetic
one.\cite{Tsunetsugu} On the other hand, the ferromagnetic model at
half-filling is also insulating with a Haldane type spin
gap.\cite{Shibata1} For lower band fillings, the numerical evidence
shows three distinct phases: a phase with partial ferromagnetic order
and incommensurate spin correlations, a fully saturated
ferromagnetically ordered phase and a region with phase separation
where two kinds of ground state seem to compete.  The energy scales
where the transitions take place for both models are of the order of
the Fermi energy. Finally, there is strong numerical evidence in favor
of a Luttinger Liquid behavior in the paramagnetic phase of the AFM
KLM, even for considerably large coupling constants.\cite{STU97, Ueda,
Wat00} Such phenomenology is beyond a simple RKKY versus Kondo type of
picture,\cite{Gulacsi2} as proposed by Doniach for the higher
dimensional models.\cite{Doniach} In fact, the level crossing
responsible for the quantum critical point (QCP) is related to long
wavelength modes. This is in contrast with the short wavelength spin
modes involved in the paramagnetic-antiferromagnetic transition that
the Doniach's picture envisages. As pointed out
before,\cite{Tsunetsugu, Gulacsi2} the missing element is the lowering
of the conduction electron kinetic energy with the alignment of the
localized spin as in the double exchange mechanism. Essentially, this
is the reason why the 1D FM and AFM KLM have similar phase diagrams.

Despite these successes, it would be considerably more informative
if some analytical understanding could be gained. Even though there
have been some partial successes\cite{Zachar-Emery-Kivelson,Gulacsi1,Gulacsi2,Zac01},
there is still room for improvement. Motivated by the enormous success
of renormalization group (RG) analyses in the few impurity problem,\cite{Yuval-Anderson1,Yuval-Anderson2,Yuval-Anderson3,Wilson2,BAJ88}
we set out to apply scaling ideas to the 1D lattice case as well.
However, an RG treatment of the KLM has never, to our knowledge, been
achieved. Technically, although we know how to progressively decimate
the spins or the conduction electron states, no one has devised a
way of doing both simultaneously, specially with the incorporation
of local Kondo physics.

It is the aim of this article to put forth such a decimation scheme
for one-dimensional models of spins and fermions, in particular the
anisotropic Kondo lattice model. We draw a great deal of inspiration
from the original work of Anderson, Yuval and Hamman for the single
impurity Kondo model,\cite{Yuval-Anderson1,Yuval-Anderson2,Yuval-Anderson3}
mapping the KLM into a classical Coulomb gas, which is then decimated
by standard methods.\cite{Niehnus} This task is made simpler by the
use of bosonization methods. We therefore study the stability of the
non-interacting ground state with respect to the Kondo interaction
as a function of the coupling constants. Our study reveals that there
is no {}``weak coupling'' flow in the entire parameter space. Nevertheless,
the different {}``strong coupling'' flows of the RG equations allow
us to assign the magnetic ground states that emerge, establishing
the phase diagram for both signs of the coupling constant in a unified
fashion. While our approach in part builds upon previous studies,\cite{Zachar-Emery-Kivelson,Gulacsi1,Gulacsi2,Zac01}
it also puts on a firmer basis the procedure of neglecting backward
scattering terms in the KLM away from half-filling. As another application
of our Coulomb gas treatment, we also establish the phase diagram
of the one dimensional Ising-Kondo model.\cite{Sikkema1}

In Section~\ref{sectionpartitionfunction}, we develop a path integral
formulation of the bosonized 1D KLM. The partition function is mapped
into a two-dimensional generalized classical Coulomb gas in Section~\ref{cgsec}.
In Section~\ref{sectionrgequations}, the RG equations of the Coulomb
gas are derived and solved. Their physical interpretation is given
in Section~\ref{eff-ham}, where an effective Hamiltonian for the
renormalized Coulomb gas is obtained. The phase diagram of the model
is established in Section~\ref{1dklsec}. The Ising-Kondo model is
discussed in Section~\ref{ising-kondo}, where its phase diagram is established.
We wrap up with a brief
discussion of the relation between our and previous results in Section~\ref{discconc}.
Some more technical developments can be found in the Appendices.

\section{Partition Function\label{sectionpartitionfunction}}

We start by writing the 1D KLM Hamiltonian. The traditional Kondo
model is isotropic in spin space. Since we are going to use Abelian
bosonization, it is natural to break the \( SU\left( 2\right)  \)
symmetry down to \( U\left( 1\right)  \)

\begin{eqnarray}
H & = & -t\sum _{j,\sigma }\left( c^{\dagger }_{j+1\sigma }c^{\phantom {\dagger }}_{j\sigma }+h.c.\right) \nonumber \\
 & + & J_{\perp }\left( S_{j}^{x}s_{j}^{x}+S_{j}^{y}s_{j}^{y}\right) +J_{z}S^{z}_{j}s^{z}_{j},\label{modelodekondo} 
\end{eqnarray}
where \( c_{j\sigma } \) destroys a conduction electron in site \( j \)
with spin projection \( \sigma  \), \( \mathbf{S}_{j} \) is a localized
spin \( \frac{1}{2} \) operator and \( \mathbf{s}_{j}=\frac{1}{2}\sum _{\alpha \beta }c^{\dagger }_{j\alpha }\sigma c^{\phantom {\dagger }}_{j\beta } \),
the conduction electron spin density. We will focus on the continuum,
long-distance limit of the conduction electrons. In this case, one
can linearize the dispersion around the non-interacting \( \left( J_{z,\perp }=0\right)  \)
Fermi points \( \pm k_{F} \), where \( k_{F}a=\frac{\pi }{2}n \)
and \( n \) is the conduction electron number density, and take the
continuous limit of the fermionic operators in terms of left and right
moving field operators\cite{Zachar-Emery-Kivelson}

\begin{eqnarray*}
H & = & -iv_{F}\sum _{\sigma }\int dx\left( \psi ^{\dagger }_{R,\sigma }\partial _{x}\psi ^{\phantom {\dagger }}_{R,\sigma }-\psi ^{\dagger }_{L,\sigma }\partial _{x}\psi ^{\phantom {\dagger }}_{L,\sigma }\right) \\
 & + & \frac{aJ_{z}}{2}\sum _{j}\sum _{\alpha ,\beta ,s}\psi _{\alpha ,s}^{\dagger }\left( j\right) \psi ^{\phantom {\dagger }}_{\beta ,s}\left( j\right) \sigma ^{z}_{s,s}S^{z}\left( j\right) \\
 & + & \frac{aJ_{\bot }}{2}\sum _{i}\sum _{\alpha ,\beta ,s,s'}\psi _{\alpha ,s}^{\dagger }\left( j\right) \psi ^{\phantom {\dagger }}_{\beta ,s'}\left( j\right) \sigma ^{\eta }_{s,s'}S^{\eta }\left( j\right) ,
\end{eqnarray*}
where \( \alpha ,\beta =L\, \, \mathrm{or}\, \, R \), \( v_{F}=2t\sin k_{F}a \)
is the Fermi velocity and \( a \) is the lattice spacing. The field
operators can now be bosonized \emph{with the inclusion of the so-called
Klein factors} in usual notation\emph{\cite{Delft-Schoeller}\begin{eqnarray*}
\psi _{R,\sigma }\left( x\right)  & = & \frac{F_{R,\sigma }}{\sqrt{2\pi \alpha }}e^{i\sqrt{\pi }\left[ \phi _{\sigma }\left( x\right) -\theta _{\sigma }\left( x\right) \right] +ik_{F}x},\\
\psi _{L,\sigma }\left( x\right)  & = & \frac{F_{L,\sigma }}{\sqrt{2\pi \alpha }}e^{-i\sqrt{\pi }\left[ \phi _{\sigma }\left( x\right) +\theta _{\sigma }\left( x\right) \right] -ik_{F}x}.
\end{eqnarray*}
}One can then rewrite the Hamiltonian in terms of the charge and spin
fields\begin{eqnarray*}
\phi _{c,s}\left( x\right)  & = & \left( \phi _{\uparrow }\left( x\right) \pm \phi _{\downarrow }\left( x\right) \right) /\sqrt{2},\\
\theta _{c,s}\left( x\right)  & = & \left( \theta _{\uparrow }\left( x\right) \pm \theta _{\downarrow }\left( x\right) \right) /\sqrt{2},
\end{eqnarray*}
 as

\begin{equation}
\label{bosonham}
H=H_{0}+H_{z}^{f}+H_{\perp }^{f}+H_{z}^{b}+H_{\perp }^{b},
\end{equation}
with:

\begin{widetext}\begin{subequations}\label{hamitems}\begin{eqnarray}
H_{0} & = & \frac{v_{F}}{2}\sum _{\nu =s,c}\int dx\left( \partial _{x}\phi _{\nu }\right) ^{2}+\left( \partial _{x}\theta _{\nu }\right) ^{2}\label{freeham} \\
H^{f}_{z} & = & \sum _{x}\, J^{f}_{z}\sqrt{\frac{2}{\pi }g_{\sigma }}\partial _{x}\phi _{s}\left( x\right) S^{z}\left( x\right) \label{forward-z} \\
H^{f}_{\perp } & = & \sum _{x}\frac{J^{f}_{\perp }}{2\pi \alpha }e^{i\sqrt{\frac{2\pi }{g_{\sigma }}}\theta _{s}\left( x\right) }\cos \left[ \sqrt{2\pi g_{\sigma }}\phi _{s}\left( x\right) \right] S^{-}\left( x\right) +h.c.\label{forward-perp} \\
H^{b}_{z} & = & \sum _{x}\frac{2J^{b}_{z}}{\pi \alpha }\sin \left[ \sqrt{2\pi g_{\rho }}\phi _{c}\left( x\right) +2k_{F}x\right] \sin \left[ \sqrt{2\pi g_{\sigma }}\phi _{s}\left( x\right) \right] S^{z}\left( x\right) \label{backward-z} \\
H^{b}_{\perp } & = & \sum _{x}\frac{J^{b}_{\perp }}{2\pi \alpha }e^{i\sqrt{\frac{2\pi }{g_{\sigma }}}\theta _{s}\left( x\right) }\cos \left[ \sqrt{2\pi g_{\rho }}\phi _{c}\left( x\right) +2k_{F}x\right] S^{-}\left( x\right) +h.c.\label{backward-perp} 
\end{eqnarray}

\end{subequations}\end{widetext}where \( g_{\sigma }=g_{\rho }=1 \)
and \( \alpha \sim k^{-1}_{F} \). \( H_{0} \) is the free bosonic
Hamiltonian written as a function of \( \theta _{s,c} \) and \( \phi _{s,c} \).
We have introduced the new parameters \( g_{\sigma } \) and \( g_{\rho } \)
for future use. The {}``relativistic'' description enforced by us
broke the interaction term in two different components: forward-scattering,
\( H^{f} \), and back-scattering, \( H^{b} \). They involve the
spin current and the \( 2k_{F} \) component of the magnetization
of the non-interacting electron gas, respectively,\cite{Aff90, Gogolin}
since it is well known that the main contributions to the spin susceptibility
of the electron gas at low frequencies come from \( q\sim 0 \) and
\( q\sim 2k_{F} \). For further generalization, we will consider
the 4 parameters \( J^{f,b}_{\perp ,z} \), as independent.\cite{Zachar-Emery-Kivelson}
It is important to note that the cosines and sines of the bosonic
fields in Eqs.~(\ref{hamitems}) are just a short form notation.
Forward and backward Klein factors do not have common eigenvectors.\cite{Senechal}
Thus, we shall not neglect their contribution to the simultaneous
treatment of \( H^{f} \) and \( H^{b} \). 

\begin{figure}
{\centering \resizebox*{1.5in}{!}{\includegraphics{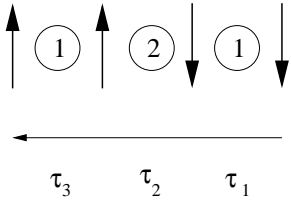}} \par}

\caption{\label{historyofaspinexample}Example of a spin history. 1 stands
for \protect\( S_z\protect \) operators (Eq.~(\ref{nokinkterms}))
and \protect\( 2\protect \) for \protect\( S_\perp \protect \)ones
(Eq.~(\ref{kinkterms})). }
\end{figure}

A quantum system of dimension \( d \) can be mapped into a classical
system of dimension \( d+1 \).\cite{Kogut1, Baxter-1} The single
impurity Kondo problem has effective dimension \( d=0 \). The work
of Anderson, Yuval, and Hamman\cite{Yuval-Anderson1,Yuval-Anderson2,Yuval-Anderson3}
showed that it can be mapped into a \( d=1 \) classical Coulomb gas,
where the extra dimension is the imaginary time.\cite{Negele, Gogolin}
We will extend this idea and map the 1D KLM into a \( d=2 \) classical
problem. As usual, the starting point is the partition function:

\begin{equation}
\label{partitonfunction}
Z=Tr\left[ e^{-\beta \left( H_{0}+H^{f}+H^{b}\right) }\right] .
\end{equation}
We will rescale the Hamiltonian and \( \beta  \) by the Fermi velocity.
This introduces the dimensionless coupling constants \( \tilde{J}_{z,\bot }=\frac{aJ_{z,\bot }}{v_{F}} \)
as well as \( \tilde{\beta }=v_{F}\beta  \). Following the standard
prescription,\cite{Fradkin, Negele} we divide \( \tilde{\beta } \)
in infinitesimal parts, 

\[
Z=Tr\left[ \prod _{j}e^{-\delta \tau H}\right] .\]
 In order to proceed to a path integral formulation we choose the
\( S^{z} \) basis for the local moments and the coherent states for
the bosonic fields.\cite{Fradkin, Negele} The next step is to introduce
an identity resolution between each exponential in the product \[
Z=\prod _{j}\left\langle \zeta \left( \tau _{j}\right) \right| e^{-\delta \tau H}\left| \zeta \left( \tau _{j}\right) \right\rangle ,\]
 where we used \( \left| \zeta \right\rangle  \) to denote a general
vector in the basis. We now expand the exponentials in powers of \( \delta \tau  \),

\[
Z=\sum _{n=0}\frac{1}{n!}\left( \delta \tau \right) ^{n}\left\langle \zeta \left( j\right) \right| H^{n}\left| \zeta \left( \tau _{j+1}\right) \right\rangle .\]
 There are two possible spin configurations for a given pair of consecutive
instants along the imaginary time direction:

\begin{enumerate}
\item if there is no spin flip between them, the only contributing terms
are from the \( z \) components of the Hamiltonian (\ref{forward-z})
and (\ref{backward-z})\begin{eqnarray}
 &  & \left\langle \vec{s}\left( x,\tau +\delta \tau \right) \right| S^{z}\left( x\right) \partial _{x}\phi _{s}\left( x\right) \left| \vec{s}\left( x,\tau \right) \right\rangle ;\nonumber \\
 &  & \left\langle \vec{s}\left( x,\tau +\delta \tau \right) \right| S^{z}\left( x\right) \sin \left[ \sqrt{2\pi g_{\rho }}\phi _{c}\left( x\right) +2k_{F}x\right] \nonumber \\
 &  & \sin \left[ \sqrt{2\pi g_{\sigma }}\phi _{s}\left( x\right) \right] \left| \vec{s}\left( x,\tau \right) \right\rangle ;\label{nokinkterms} 
\end{eqnarray}

\item if there is a spin flip between these 2 instants, then only (\ref{forward-perp})
and (\ref{backward-perp}) contribute\begin{eqnarray}
 &  & \left\langle \vec{s}\left( x,\tau +\delta \tau \right) \right| e^{i\sqrt{\frac{2\pi }{g_{\sigma }}}\theta _{s}\left( x\right) }\cos \left[ \sqrt{2\pi g_{\sigma }}\phi _{s}\left( x\right) \right] \nonumber \\
 &  & S^{-}\left( x\right) +\mathrm{h}.\mathrm{c}.\left| \vec{s}\left( x,\tau \right) \right\rangle ;\nonumber \\
 &  & \left\langle \vec{s}\left( x,\tau +\delta \tau \right) \right| e^{i\sqrt{\frac{2\pi }{g_{\sigma }}}\theta _{s}\left( x\right) }\cos \left[ \sqrt{2\pi g_{\rho }}\phi _{c}\left( x\right) +2k_{F}x\right] \nonumber \\
 &  & S^{-}\left( x\right) +\mathrm{h}.\mathrm{c}.\left| \vec{s}\left( x,\tau \right) \right\rangle .\label{kinkterms} 
\end{eqnarray}

\end{enumerate}
The two possible processes above are illustrated in Fig.~\ref{historyofaspinexample}.
Several bosonic operators can fit inside the example given in that
figure, for example

\begin{widetext}\begin{eqnarray*}
 &  & \frac{\tilde{J}^{b}_{z}\tilde{J}^{f}_{\perp }}{2}\left( \sqrt{\frac{2}{\pi }g_{\sigma }}\tilde{J}^{f}_{z}\right) F_{L\uparrow }^{\dagger }F_{R\uparrow }F_{R\downarrow }^{\dagger }F_{R\uparrow }S^{z}\left( x,\tau _{3}\right) S^{+}\left( x,\tau _{2}\right) S^{z}\left( x,\tau _{1}\right) \times \\
 &  & e^{i\sqrt{2\pi g_{\sigma }}\phi _{s}\left( x,\tau _{3}\right) +i\sqrt{2\pi g_{\rho }}\phi _{c}\left( x,\tau _{3}\right) }e^{i\sqrt{\frac{2\pi }{g_{\rho }}}\theta _{s}\left( x,\tau _{2}\right) -i\sqrt{2\pi g_{\sigma }}\phi _{s}\left( x,\tau _{2}\right) +2ik_{F}x}\partial _{x}\phi _{s}\left( x,\tau _{1}\right) ,
\end{eqnarray*}

\end{widetext}where we regrouped terms to separate Klein factors,
local spin operators and bosonic fields.

The lattice parameter in the Euclidean-time direction is set by the
bosonic cut-off: \( \delta \tau \cong 2\pi \alpha  \). Keeping the
leading order terms, we write the partition function as the sum over
all \( d=2 \) Ising spin configurations of the localized spins, Klein
factors and a functional integral over the bosonic variables. Since
there are different types of bosonic exponentials ({}``vertex operators''),
coming from the different interaction terms in Eq.~(\ref{bosonham}),
we now introduce new Ising variables which we will call {}``charges''
in order to do the bookkeeping. They give the sign of the corresponding
bosonic field in the accompanying exponential according to the following
scheme:

\begin{enumerate}
\item \( m\left( x,\tau \right) =S^{z}\left( x,\tau +\delta \tau \right) -S^{z}\left( x,\tau \right) =\pm 1 \)
gives the sign of \( \theta _{s}\left( x,\tau \right)  \) (Eqs.~(\ref{forward-perp})
and (\ref{backward-perp}))
\item \( e\left( x,\tau \right) =\pm 1 \) gives the sign of \( \phi _{s}\left( x,\tau \right)  \)
(Eqs.~(\ref{forward-perp}) and (\ref{backward-z}))
\item \( c\left( x,\tau \right) =\pm 1 \) gives the sign of \( \phi _{c}\left( x,\tau \right)  \)
(Eqs.~(\ref{backward-z}) and (\ref{backward-perp}))
\end{enumerate}
Note that only \( m\left( x,\tau \right)  \) is always tied to a
localized spin flip process, its value giving both the sign of the
\( \theta _{s} \) coefficient and the change in \( S^{z} \). With
this notation, each point in the Euclidean {}``space-time'' is labeled
by a triad of values \( \left( m,e,c\right)  \). We call a {}``particle''
a point where \( \left( m,e,c\right) \neq \left( 0,0,0\right)  \).
Each kind of particle matches a certain incident in the history of
a spin. From Eqs.~\ref{hamitems}, we can read off the existence
of three breeds of particles, each one with its respective fugacity.
Table~\ref{initialparticlestable} summarizes the notation that we
will use.

\begin{table}
{\centering \begin{tabular}{|c|c|c|}
\hline 
fugacity&
\( \left( m,e,c\right)  \)&
Number of particles\\
\hline
\hline 
\( y_{1}=\tilde{J}^{f}_{\perp }/2 \)&
\( \left( \pm 1,\pm 1,0\right)  \)&
\( N_{1} \)\\
\hline 
\( y_{2}=\tilde{J}^{b}_{\perp }/2 \)&
\( \left( \pm 1,0,\pm 1\right)  \)&
\( N_{2} \)\\
\hline 
\( y_{3}=\tilde{J}^{b}_{z}/2 \)&
\( \left( 0,\pm 1,\pm 1\right)  \)&
\( N_{3} \)\\
\hline
\end{tabular}\par}

\caption{\label{initialparticlestable}Particles in the 1D KLM and their charges.}
\end{table}

Denoting \( \eta ^{l}_{j} \) as the space-time position of particle
\( j \) of type \( l=\left\{ 1,2,3\right\}  \) and \( {\cal D}\eta =\prod ^{3}_{l=1}\prod ^{N_{l}}_{j=1}d\eta _{j}^{l} \)
, we can write the partition function as:

\begin{widetext}

\begin{eqnarray}
Z & = & \sum _{\left\{ \sigma \right\} }\sum _{N_{1},N_{2},N_{3}=0}^{\infty }\sum _{\left\{ m,e,c\right\} }\int {\cal D}\phi _{s,c}{\cal D}\theta _{s,c}{\cal D}\eta \, \frac{y_{1}^{N_{1}}y_{2}^{N_{2}}y_{3}^{N_{3}}}{N_{1}!N_{2}!N_{3}!}\, \left( \textrm{Klein factors}\right) \, \left[ \prod _{j=1}^{N_{3}}\sigma ^{z}\left( \eta _{j}^{3}\right) \right] \nonumber \\
 & \exp  & \left\{ -S_{0}-\tilde{J}^{f}_{z}\sqrt{\frac{2}{\pi }g_{\sigma }}\sum _{x}\int d\tau \partial _{x}\phi _{s}\left( x,\tau \right) S^{z}\left( x,\tau \right) +i\sqrt{\frac{2\pi }{g_{\sigma }}}\sum _{l=1,2}\sum _{\eta ^{l}_{j}}m\left( \eta ^{l}_{j}\right) \theta _{s}\left( \eta _{j}^{l}\right) \right. \nonumber \\
 & + & \left. i\sqrt{2\pi g_{\sigma }}\sum _{l=1,3}\sum _{\eta ^{l}_{j}}e\left( \eta _{j}^{l}\right) \phi _{s}\left( \eta _{j}^{l}\right) +\sqrt{2\pi g_{\rho }}\sum _{l=2,3}\sum _{\eta _{j}^{l}}\left[ c\left( \eta _{j}^{l}\right) \phi _{c}\left( \eta _{j}^{l}\right) +2ik_{F}c\left( \eta _{j}^{l}\right) x^{l}_{j}\right] \right\} ,\label{partitionfunctionwithbosons} 
\end{eqnarray}

\end{widetext}where \( \left\{ \sigma \right\}  \) stands for all
Ising spin configurations, \( \left\{ m,e,c\right\}  \) represents
all possible sets of \( \sum _{i}N_{i} \) particles, \( x^{l}_{j} \)
is the space coordinate of particle \( \eta ^{l}_{j} \) and \( S_{0} \)
is the free Gaussian bosonic action in the variables \( \phi _{s,c} \)
and \( \theta _{s,c} \)\cite{Gogolin}. There are several restrictions
over \( \left\{ m,e,c\right\}  \), usually called neutrality conditions
in the bosonization\cite{Gogolin} and Coulomb Gas\cite{Niehnus}
literatures. In the 1D KLM they are more stringent than usual, ensuring
the compatibility of the sums over \( \left\{ \sigma \right\}  \)
and \( \left\{ m,e,c\right\}  \) in Eq~(\ref{partitionfunctionwithbosons}).
Therefore, we will call them strong neutrality conditions (see Appendix
\ref{neutralityapendix} for the derivation of these conditions).
They are:

\begin{enumerate}
\item for each space coordinate the \( m \) charges must be neutral, \( \sum _{i}m\left( x_{fixed},\tau _{i}\right) =0 \),
\item the total \( e \) charge must be neutral, \( \sum _{i}e\left( x_{i},\tau _{i}\right) =0 \),
\item for each space coordinate the total charge \( c \) must be an even
integer \( \sum _{i}c\left( x_{fixed},\tau _{i}\right) =2n,\, n\in \mathbb Z \);
and the total charge in the entire space-time must be zero, \( \sum _{i}c\left( x_{i},\tau _{i}\right) =0 \).
\end{enumerate}
We can immediately see two consequences of these conditions. The most
obvious is that the sign of \( \tilde{J}^{f,b}_{\perp } \) is irrelevant
since, from condition 1, the total number of spin flips in the time
direction \( N_{1}+N_{2} \) is even. The other consequence is more
subtle and more surprising: the complete cancellation of the Klein
factors and the product of \( \sigma ^{z}\left( \eta _{j}^{3}\right)  \)
in Eq.~(\ref{partitionfunctionwithbosons}),

\[
\left( \textrm{Klein factors}\right) \prod _{j=1}^{N_{3}}\sigma ^{z}\left( \eta ^{3}_{j}\right) =1.\]
This result plays a central role in the renormalization group treatment
of a single Kondo impurity in a LL by Lee and Toner.\cite{Lee} Moreover,
it leads to:

\begin{equation}
\label{comensurabilitycond}
2k_{F}\sum _{i}c\left( i\right) x_{i}=\left\{ \begin{array}{l}
0,\\
\\
4k_{F}aI,\, I\in \mathbb Z
\end{array}\right. ,
\end{equation}
in each contribution to the partition function. The \( 2k_{F} \)
terms appear whenever there are particles of type \( 2 \) and \( 3 \)
(see the definition of the charge \( c \) and Eqs.~(\ref{backward-z})
and (\ref{backward-perp})). Due to their oscillatory nature, configurations
with these particles will be strongly suppressed in the statistical
sum and the corresponding terms (with fugacities \( y_{2}=\tilde{J}_{\perp }^{b}/2 \)
and \( y_{3}=\tilde{J}_{z}^{b}/2 \)) will be irrelevant in the RG
sense. This irrelevance criterion is precisely the same as the one
used from neglecting Umklapp scattering away from half-filling in
models like the Hubbard model.\cite{Voit} However, we stress that
the situation here is far less trivial than in the Hubbard model,
since we have both fermions and spins, and the latter have no independent
dynamics, thus hindering a rigorous analysis (an important exception
to this is the Heisenberg-Kondo model\cite{Sikkema2}). In our treatment,
on the other hand, spins and fermions are treated on the same footing
and lose their independent identity. After the mapping to a Coulomb
gas, the irrelevance criterion becomes identical to other models where
its applicability is firmly based. We have thus established a more
rigorous basis for neglecting the backward scattering terms in the
1D KLM, as has been done by Zachar, Kivelson and Emery.\cite{Zachar-Emery-Kivelson}

This situation changes when the conduction band is at half filling.
In this case, \( 4k_{F}a=2\pi  \) and these terms disappear from
the effective action, making all particles equally probable. This
commensurability condition is similar to the one for the Umklapp term
in the Hubbard model.\cite{Voit} It is interesting to note that only
the combination \( 4k_{F}a \) appears in our formulation. Since we
bosonized the non-interacting conduction electron sea, we must use
\( k_{F}a=\pi n/2 \), leading to \( 4k_{F}a=2\pi n \). Even if for
some reason a large Fermi surface should be considered,\cite{YOA97}
this would not change any of our results since for a large Fermi surface
\( 4k_{F}^{*}a=2\pi \left( n+1\right) =4k_{F}a\, \mathrm{mod}\, 2\pi  \).

\section{Coulomb Gas\label{cgsec}}

The bosonic fields in Eq.~(\ref{partitionfunctionwithbosons}) can
now be integrated out, partially summing the partition function. The
result can be understood as an effective action for the spins and
the variables \( \left( m,e,c\right)  \)

\begin{widetext}

\begin{eqnarray}
S_{eff} & = & \left( \frac{\tilde{J}_{z}^{f}}{\pi }\right) ^{2}\sum _{x_{1}>x_{2}}\int _{\tau _{1}>\tau _{2}}d\tau _{1}d\tau _{2}\frac{\cos \left( 2\varphi _{12}\right) }{r_{12}^{2}}S^{z}\left( 1\right) S^{z}\left( 2\right) \nonumber \\
 & + & \frac{\tilde{J}_{z}^{f}}{\pi }\sum _{n}\sum _{x}\int d\tau \frac{\exp \left( ie\left( n\right) \varphi \right) }{r}S^{z}\left( x,\tau \right) \nonumber \\
 & + & \sum _{n>p}\frac{\ln z_{np}}{2}\left( m\left( n\right) +e\left( n\right) \right) \left( m\left( p\right) +e\left( p\right) \right) +\frac{\ln \bar{z}_{np}}{2}\left( m\left( n\right) -e\left( n\right) \right) \left( m\left( p\right) -e\left( p\right) \right) \nonumber \\
 & + & \ln \left| r_{np}\right| c\left( n\right) c\left( p\right) +2ik_{F}\sum _{n}c\left( n\right) x_{l}+\mathrm{short}\, \mathrm{range}\, \mathrm{interactions},\label{effectiveaction} 
\end{eqnarray}

\end{widetext}where\begin{eqnarray}
x_{jk} & = & x_{k}-x_{j},\nonumber \\
\tau _{jk} & = & \tau _{k}-\tau _{j},\nonumber \\
z_{jk} & = & x_{jk}+i\tau _{jk}=r_{jk}e^{i\varphi _{jk}}.\label{zjk} 
\end{eqnarray}

In addition to the long range universal interactions, this procedure
also gives rise to short-ranged terms that are cutoff dependent.\cite{Wiegmann}
These are similar to those found by Honner and Gulácsi\cite{Gulacsi1, Gulacsi2}
by smoothing the bosonic commutation relations. In contrast to their
treatment, though, here they are a manifestation of the bosonic field
dynamics. Following Zachar \emph{et. al},\cite{Zachar-Emery-Kivelson}
we will focus on the universal long range part of the action and neglect
these terms.

Upon integrating by parts in imaginary time, spin time derivatives
become the charges we denote by \( m \). Finally, we can rewrite
all long range terms in the form of a generalized CG action in two-dimensional
Euclidean space\cite{Niehnus} as

\begin{equation}
Z=\sum _{N_{1},N_{2},N_{3}=0}^{\infty }\sum _{\left\{ m,e,c\right\} }\int D\eta \frac{y_{1}^{N_{1}}y_{2}^{N_{2}}y_{3}^{N_{3}}}{N_{1}!N_{2}!N_{3}!}\exp \left\{ S_{eff}\right\} ,
\end{equation}
with\begin{eqnarray}
S_{eff} & = & \frac{1}{2}\sum _{i\neq j}\left\{ \frac{\kappa ^{2}}{g_{\sigma }}\ln \left| r_{ij}\right| m\left( \eta _{i}\right) m\left( \eta _{j}\right) \right. \nonumber \\
 & + & g_{\sigma }\ln \left| r_{ij}\right| e\left( \eta _{i}\right) e\left( \eta _{j}\right) +g_{\rho }\ln \left| r_{ij}\right| c\left( \eta _{i}\right) c\left( \eta _{j}\right) \nonumber \\
 & - & \left. i\kappa \varphi _{ij}\left[ e\left( \eta _{i}\right) m\left( \eta _{j}\right) +m\left( \eta _{i}\right) e\left( \eta _{j}\right) \right]\vphantom{\frac{\kappa ^{2}}{g_{\sigma }}}\right\} \nonumber \\
 & + & 2ik_{F}\sum _{i}c\left( \eta _{i}\right) x_{i},\label{coulombgas} 
\end{eqnarray}
 where \( \kappa =1-\tilde{J}^{f}_{z}/\pi  \). In the above effective
action we have dropped the superindex indicating the particle type
in order to unclutter the notation. It is now unnecessary as the dependence
with the history of a spin has disappeared.

In most other similar CG mappings, the coefficient of the term in
\( \varphi _{ij} \) is an integer and goes by the name of conformal
spin.\cite{Gogolin} Then, the ambiguity of \( 2\pi I,\, I\in \mathbb Z \)
in the angle is irrelevant. In this case, however, \( \kappa  \)
can assume non-integer values. What guarantees that the theory is
actually well defined is the strong neutrality condition \( 1 \),
which leads to a cancellation of the Riemann surface index \( I \).

The integration by parts that we performed is equivalent to applying
the duality relation \( \partial _{x}\phi _{s}=i\partial _{\tau }\theta _{s} \)
to Eq.~(\ref{partitionfunctionwithbosons}), integrating by parts
and then tracing the bosonic fields. Alternatively, at the Hamiltonian
level, it is also equivalent to applying the rotation\cite{Zachar-Emery-Kivelson}

\begin{equation}
\label{ZKErotation}
U=e^{i\sqrt{\frac{2}{\pi }}J^{f}_{z}\sum _{x}\theta _{s}\left( x\right) S^{z}\left( x\right) },
\end{equation}
to Eq~(\ref{bosonham}) before going to a path integral and tracing
out the bosons. Hence, there is a strong link between our CG formulation
and previous results in the literature.\cite{Zachar-Emery-Kivelson,Gulacsi1,Gulacsi2}
More importantly, the interpretation of our results should be understood
in this rotated basis that mixes spins and bosons.

The effective action in Eq.~(\ref{coulombgas}) can be viewed as
describing the electrostatic and magnetostatic energy of singly charged
particles with both electric and magnetic monopoles. These satisfy
electric-magnetic duality in the sense that the action is invariant
under the exchange \( e\rightleftharpoons m \) and \( g_{\sigma }\rightleftharpoons \frac{\kappa ^{2}}{g_{\sigma }} \),
while \( \kappa  \) is unchanged. This is analogous to the Dirac
relation between electric and magnetic monopoles. Furthermore, these
particles possess a third electric-like charge (\( c) \), unrelated
to the two previous ones. The partition function sum now has been
reduced to considering all particle configurations, blending spins
and bosons in this Coulomb gas representation, where we have particles
plus neutrality conditions.

A partially traced partition function allows us to link problems that
are originally quite distinct. For example, the only difference between
the CG's of the single impurity Kondo problem and the problem of tunneling
though an impurity in a Luttinger liquid\cite{KaneFisher} are the
neutrality conditions. Analogously, the two-channel Kondo problem\cite{Furusaki_Matveev,Hangmo_Kane}
and the double barrier tunneling\cite{KaneFisher} can be mapped into
each other with the same neutrality conditions. The KLM also has an
unsuspicious counterpart in the literature: two weakly coupled spinless
Luttinger liquids.\cite{Kusmartsev, Nersesyan, Gogolin} The tunneling
from one LL to the other is analogous to a spin flip process that
scatters a boson from an up spin band to a down one and vice-versa.
In particular, the two problems give the same effective action (with
different neutrality conditions) if we disregard the backward-scattering
terms in Eq.~(\ref{bosonham}) and consider the anisotropic case
\( \kappa =1 \) (\( J^{f}_{z}=0 \)). 

In the following section, we will derive the Coulomb gas renormalization
group equations following closely the review by Nienhuis.\cite{Niehnus}
As expected, the procedure strongly resembles the renormalization
group analysis of the tunneling between 2 LL's.\cite{GS88, Yakovenko1, Kusmartsev, Nersesyan, KR94, FN01, Gogolin}
The Coulomb couplings \( g_{\sigma ,\rho } \) are equal to \( 1 \)
for non-interacting conduction electrons. However, the same RG equations
will apply to the case of conduction electrons with an SU(2) non-invariant
forward scattering interaction. In this case, the initial values of
\( g_{\sigma ,\rho } \) are the corresponding Luttinger liquid parameters.\cite{Delft-Schoeller}
We will not dwell upon this case here, but its phase diagram is analogous
to the one we will derive below.

\section{Renormalization Group Equations\label{sectionrgequations}}

The philosophy of the renormalization group is to sum the partition
function by infinitesimal steps and find recursive equations for the
coupling constants while keeping the same form of the effective action.
In a Coulomb gas each step corresponds to three distinct procedures:
length rescaling, particle fusion and particle annihilation.\cite{Niehnus}
In order to implement these procedures all the particle fugacities
must be small and we are forced to impose \( J_{z}^{b} \) and \( J^{f,b}_{\perp }\ll t \).

\begin{figure}
{\centering \resizebox*{3in}{!}{\includegraphics{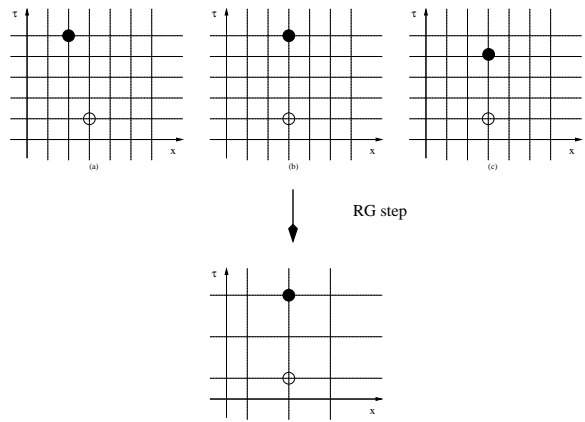}} \par}

\caption{\label{figreescala}Length rescaling in the CG. Originally distinct
charge configurations are identified at the new scale.}
\end{figure}

The first step consists of integrating large wavelength modes and
then rescaling parameters so as to reconstruct the original action
form.\cite{Shankar} This corresponds to the overall length rescaling:

\begin{equation}
\label{rescale}
r_{ij}=\frac{\bar{r}_{ij}}{1-d\ell }
\end{equation}
in the action and the partition function measure \( D\eta  \). Upon
rescaling we lose the ability to distinguish certain previously distinct
charge configurations as exemplified in Fig.~\ref{figreescala}.
Applying Eq.~(\ref{rescale}) to the effective action Eq.~(\ref{coulombgas})
and expanding the logarithm for \( d\ell \ll 1 \), we obtain:

\begin{eqnarray*}
S_{eff} & = & \bar{S}_{eff}+\frac{1}{2}\sum _{i\neq j}\left[ \frac{\kappa ^{2}}{g_{\sigma }}m\left( \eta _{i}\right) m\left( \eta _{j}\right) \right. \\
 & + & \left. g_{\sigma }e\left( \eta _{i}\right) e\left( \eta _{j}\right) +g_{\rho }c\left( \eta _{i}\right) c\left( \eta _{j}\right) \right] d\ell .
\end{eqnarray*}
The neutrality conditions can be used to rewrite the last term as
a single sum over sites:

\begin{eqnarray}
\bar{S}_{eff} & = & S_{eff}+\frac{1}{2}\sum _{i}\left[ \frac{\kappa ^{2}}{g_{\sigma }}m\left( \eta _{i}\right) ^{2}\right. \nonumber \\
 & + & \left. g_{\sigma }e\left( \eta _{i}\right) ^{2}+g_{\rho }c\left( \eta _{i}\right) ^{2}\right] d\ell .\label{rescale1} 
\end{eqnarray}
The integral over \( \eta  \) is the sum over all possible particle
positions, and its rescaling leads to:

\begin{equation}
\label{rescale2}
d\eta _{i}=\frac{d\bar{\eta }_{i}}{\left( 1-d\ell \right) ^{d}}.
\end{equation}

\begin{table}
{\centering \begin{tabular}{|c|c|c|}
\hline 
fugacity&
\( \left( m,e,c\right)  \)&
Number of particles\\
\hline
\hline 
\( \tilde{G}=0 \)&
\( \left( \pm 2,0,0\right)  \)&
\( N_{4} \)\\
\hline 
\( G=0 \)&
\( \left( 0,\pm 2,0\right)  \)&
\( N_{5} \)\\
\hline 
\( \Gamma =0 \)&
\( \left( 0,0,\pm 2\right)  \)&
\( N_{6} \)\\
\hline
\end{tabular}\par}

\caption{\label{newparticlestable}New particles created upon rescaling and
their charges.}
\end{table}

We have left the dimension \( d \) unspecified for the following
reason. Since a particle can only exist at the space coordinate where
a spin exists, we can define two important limits in the KLM. If the
Kondo spins are separated by a distance greater than \( d\ell  \),
the sum over identical configurations is one dimensional \( \left( d=1\right)  \),
as in the single impurity case (see \( \left( b\right)  \) and \( \left( c\right)  \)
of figure \ref{figreescala}). This is the dilute limit or {}``incoherent
regime'' of the Kondo lattice, where the scaling proceeds exactly
as in the single impurity Kondo problem in a LL as found by Lee and
Toner.\cite{Lee} In contrast, when \( d\ell  \) is larger than the
distance between Kondo spins we are in the dense limit\cite{Zachar-Emery-Kivelson}
or {}``coherent regime'' of the Kondo lattice. In the latter case,
the identification of initially distinct configurations can involve
charges at different space coordinates, implying that \( d=2 \) (see
\( \left( a\right)  \) in figure \ref{figreescala}). We will focus
on this coherent regime and set \( d=2 \) from now on.

Collecting Eqs.~(\ref{rescale1}) and (\ref{rescale2}) we can express
the partition function once again as a Coulomb gas by redefining the
particle fugacities. A particle with charges \( \left( m,e,c\right)  \)
has its fugacity \( Y_{m,e,c} \) renormalized as\begin{equation}
\label{rgfugacities}
\frac{dY_{m,e,c}}{d\ell }=\left( 2-\frac{1}{2}\left( \frac{\kappa ^{2}}{g_{\sigma }}m^{2}+g_{\sigma }e^{2}+g_{\rho }c^{2}\right) \right) Y_{m,e,c}.
\end{equation}

\begin{figure}
{\centering \resizebox*{3in}{!}{\includegraphics{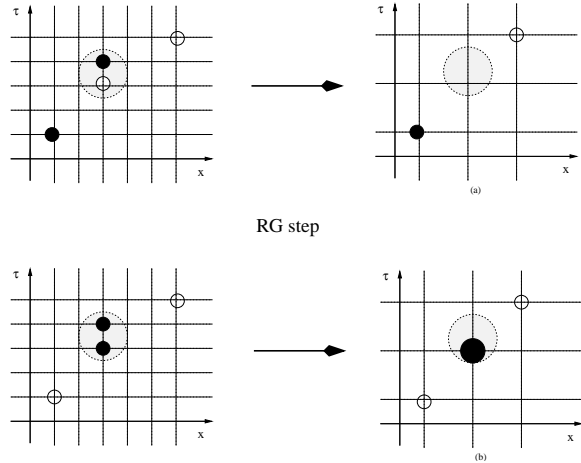}} \par}

\caption{\label{fusionandannihilationfigure}{}``Close pair'' processes in
the RG step: (a) particle annihilation and (b) particle fusion.}
\end{figure}

This equation gives the dimension of the corresponding operator and
leads to the standard relevance criteria for bosonic operators. However,
we have so far disregarded two possibilities. Suppose that a pair
of initially distinct particles is within range of the new smallest
scale. Following Anderson \emph{et al.}\cite{Yuval-Anderson3} we
call it a {}``close pair''. After the RG step we can no longer resolve
these two particles as separate entities. On the one hand, if the
particles have precisely opposite charges we have a {}``pair annihilation''
(see (a) in Fig~(\ref{fusionandannihilationfigure})). The residual
dipole polarization of this pair renormalizes the interaction among
the other particles, leading to the RG equations for \( g_{\sigma } \)
and \( g_{\rho } \). Note that \( \kappa  \) is an RG invariant.
On the other hand, if the pair is not neutral the particles are fused
into a new particle carrying the net charge (see (b) in Fig~(\ref{fusionandannihilationfigure})).
This last process may actually create particles previously absent
in the gas. There are three new kinds of particles created upon fusion
in the dense limit with initial conditions \( g_{\sigma }\sim g_{\rho }\sim 1 \).
Their charges and fugacities are listed in Tab.~\ref{newparticlestable}.
These new entities correspond to originally marginal operators that
are absent in the bare problem (their physical meaning will be discussed
in the next section). Other particles with higher charges could also
be considered, but from Eq.~(\ref{rgfugacities}) it is clear that
they are highly irrelevant and therefore can be neglected. Collecting
the annihilation and fusion terms, derived in Appendix~\ref{appendixannihilationandfusion},
and adding the dimensionality equation (\ref{rgfugacities}), we complete
the renormalization group equations. Away from half-filling, where
the backward-scattering terms are irrelevant, particles with fugacities
\( y_{2,3} \) and \( \Gamma  \) can be disregarded. Thus, only configurations
involving the fugacities \( y_{1} \), \( G \) and \( \tilde{G} \)
need to be considered. On the other hand, at half-filling all particles
from Tabs.~\ref{initialparticlestable} and \ref{newparticlestable}
should be included. This leads to the following renormalization group
equations

\begin{widetext}

\begin{itemize}
\item Away from half-filling\begin{eqnarray*}
\frac{dy_{1}}{dl} & = & \left( 2-\frac{1}{2}\left( \frac{\kappa ^{2}}{g_{\sigma }}+g_{\sigma }\right) \right) y_{1}+\frac{\sin \left( 2\pi \kappa \right) }{2\kappa }y_{1}\left( G+\tilde{G}\right) ,\\
\frac{dG}{dl} & = & 2\left( 1-g_{\sigma }\right) G+\pi y_{1}^{2},\\
\frac{d\tilde{G}}{dl} & = & 2\left( 1-\frac{\kappa ^{2}}{g_{\sigma }}\right) \tilde{G}+\pi y_{1}^{2},\\
\frac{1}{2\pi ^{2}}\frac{d\ln g_{\sigma }}{dl} & = & \frac{\sin \left( 2\pi \kappa \right) }{4\pi \kappa }\left( \frac{\kappa ^{2}}{g_{\sigma }}-g_{\sigma }\right) y_{1}^{2}+\frac{\kappa ^{2}}{g_{\sigma }}\tilde{G}^{2}-g_{\sigma }G^{2},\\
\frac{1}{2\pi ^{2}}\frac{d\ln g_{\rho }}{dl} & = & 0.
\end{eqnarray*}

\item At half-filling\begin{eqnarray*}
\frac{dy_{1}}{dl} & = & \left( 2-\frac{1}{2}\left( \frac{\kappa ^{2}}{g_{\sigma }}+g_{\sigma }\right) \right) y_{1}+\frac{\sin \left( 2\pi \kappa \right) }{2\kappa }y_{1}\left( G+\tilde{G}\right) +\pi y_{2}y_{3},\\
\frac{dy_{2}}{dl} & = & \left( 2-\frac{1}{2}\left( \frac{\kappa ^{2}}{g_{\sigma }}+g_{\rho }\right) \right) y_{2}+\frac{\sin \left( \pi \kappa \right) }{\kappa }y_{1}y_{3}+\pi y_{2}\left( \tilde{G}+\Gamma \right) ,\\
\frac{dy_{3}}{dl} & = & \left( 2-\frac{1}{2}\left( g_{\sigma }+g_{\rho }\right) \right) y_{3}+\frac{\sin \left( \pi \kappa \right) }{\kappa }y_{1}y_{2}+\pi y_{3}\left( G+\Gamma \right) ,\\
\frac{dG}{dl} & = & 2\left( 1-g_{\sigma }\right) G+\pi \left( y_{1}^{2}+y^{2}_{3}\right) ,\\
\frac{d\tilde{G}}{dl} & = & 2\left( 1-\frac{\kappa ^{2}}{g_{\sigma }}\right) \tilde{G}+\pi \left( y_{1}^{2}+y^{2}_{2}\right) ,\\
\frac{d\Gamma }{dl} & = & 2\left( 1-g_{\rho }\right) \Gamma +\pi \left( y^{2}_{2}+y^{2}_{3}\right) ,\\
\frac{1}{2\pi ^{2}}\frac{d\ln g_{\sigma }}{dl} & = & \frac{\sin \left( 2\pi \kappa \right) }{4\pi \kappa }\left( \frac{\kappa ^{2}}{g_{\sigma }}-g_{\sigma }\right) y_{1}^{2}+\frac{\kappa ^{2}}{g_{\sigma }}\left( \tilde{G}^{2}+\frac{y^{2}_{2}}{2}\right) -g_{\sigma }\left( G^{2}+\frac{y^{2}_{3}}{2}\right) ,\\
\frac{1}{2\pi ^{2}}\frac{d\ln g_{\rho }}{dl} & = & -g_{\rho }\left( \frac{y^{2}_{2}}{2}+\frac{y_{3}^{2}}{2}+\Gamma ^{2}\right) ;
\end{eqnarray*}

\end{itemize}
\end{widetext}

\begin{figure}
{\centering \subfigure[Fugacities]{\resizebox*{3in}{!}{\includegraphics{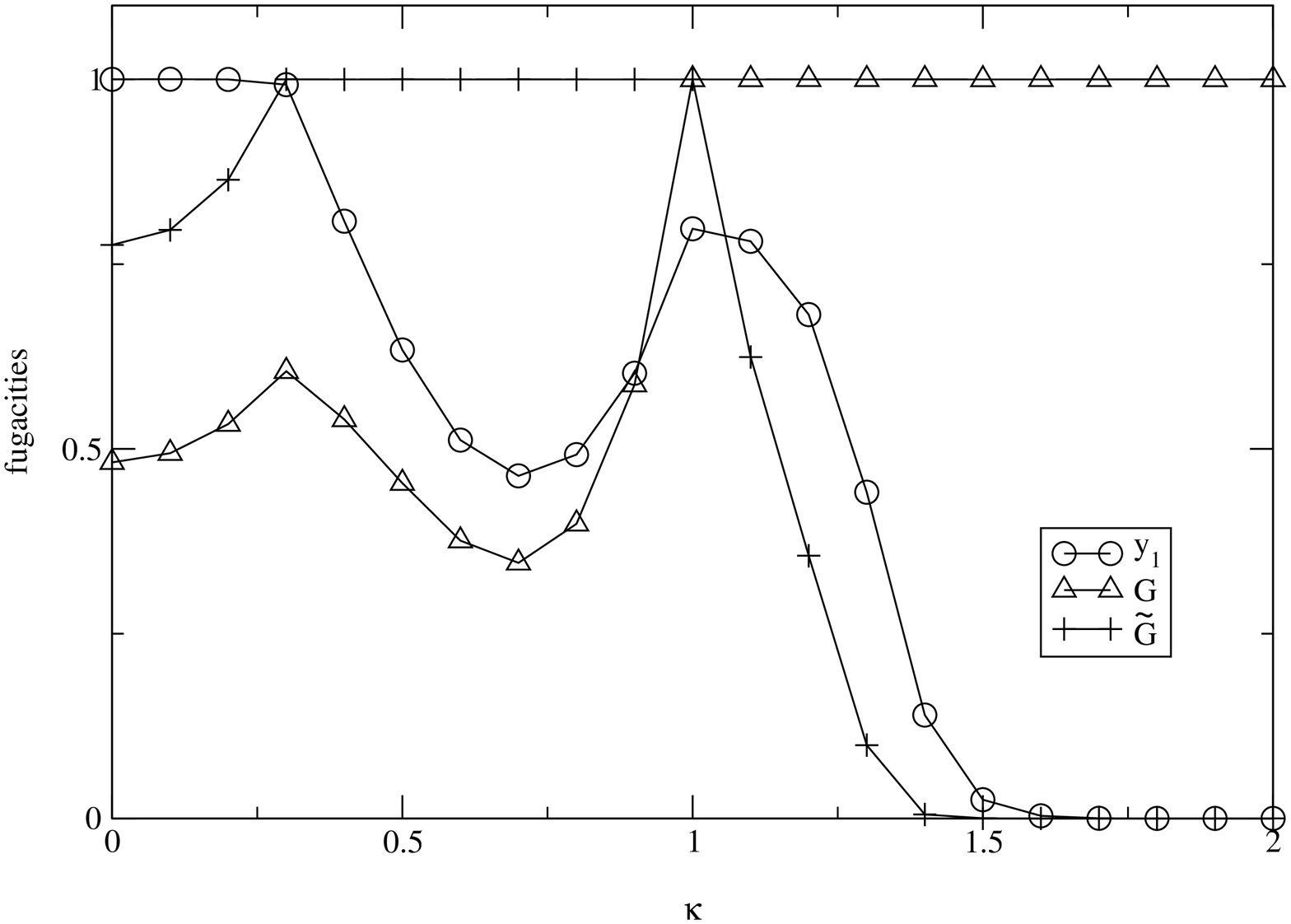}}} \par}

{\centering \subfigure[Interaction constants]{\resizebox*{3in}{!}{\includegraphics{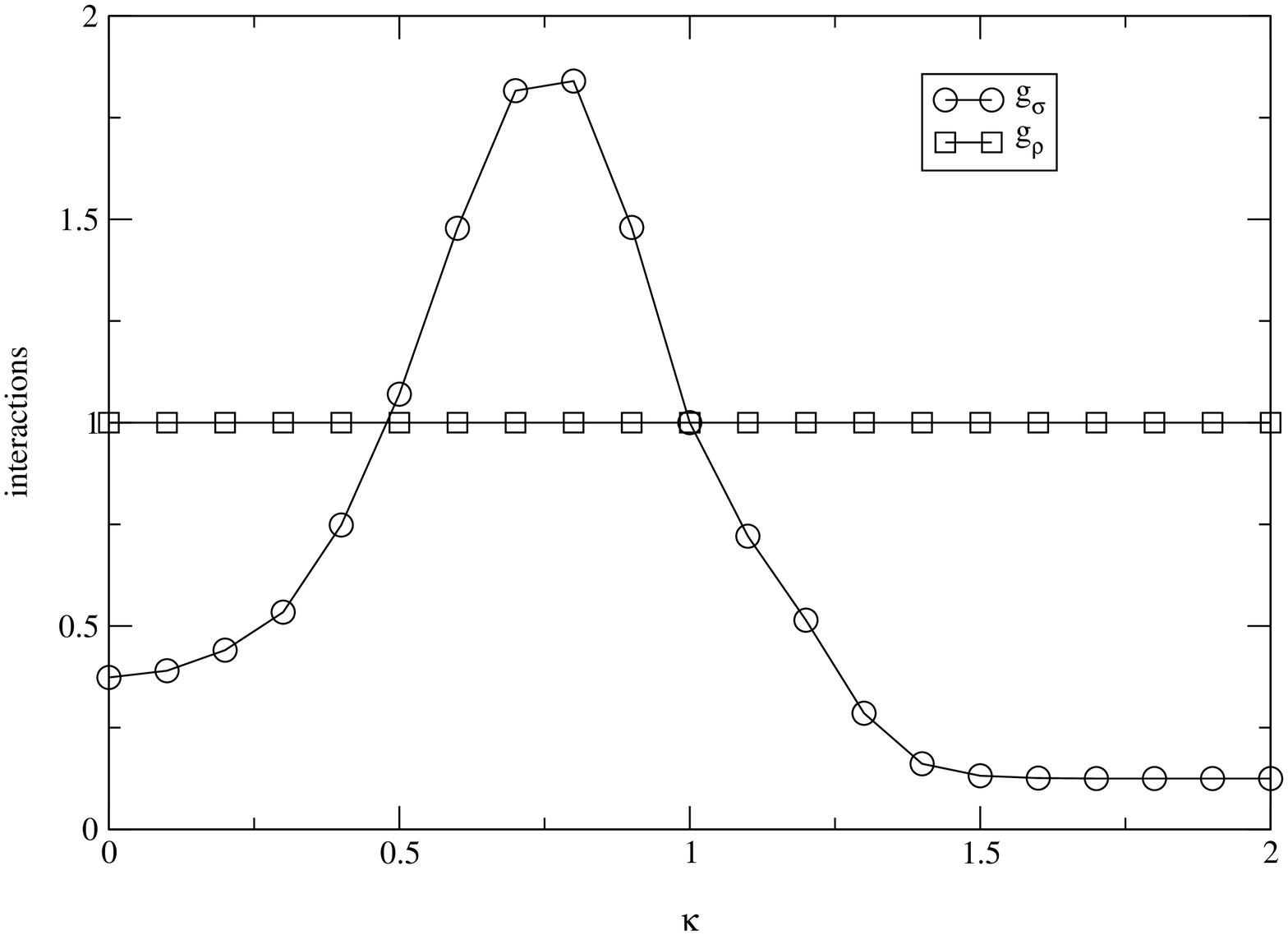}}} \par}

\caption{\label{numericalsolutionouthalf}RG flow away from half-filling as
a function of \protect\( \kappa \protect \).}
\end{figure}

\begin{figure}
{\centering \subfigure[Fugacities]{\resizebox*{3in}{!}{\includegraphics{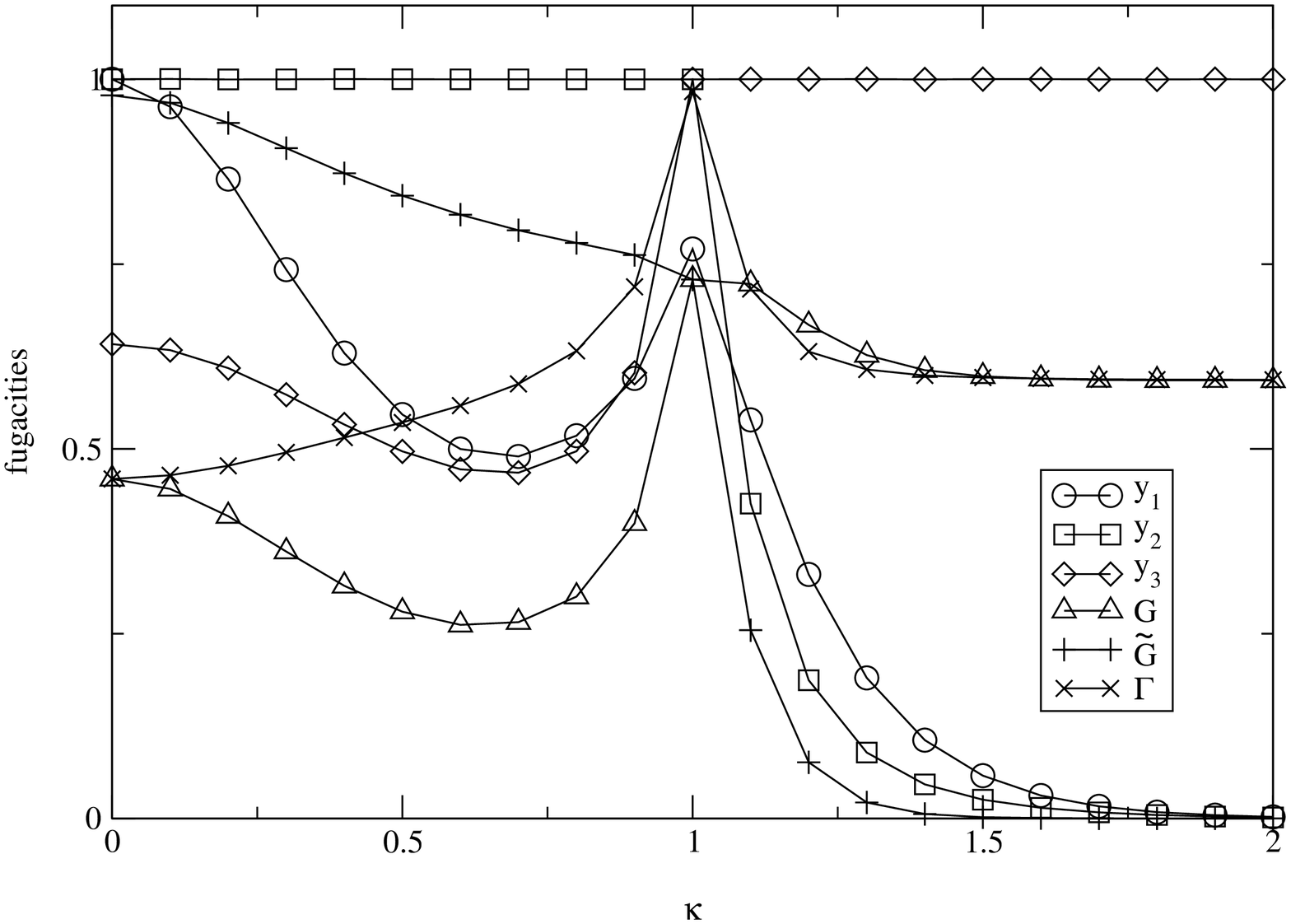}}} \par}

{\centering \subfigure[Interaction constants]{\resizebox*{3in}{!}{\includegraphics{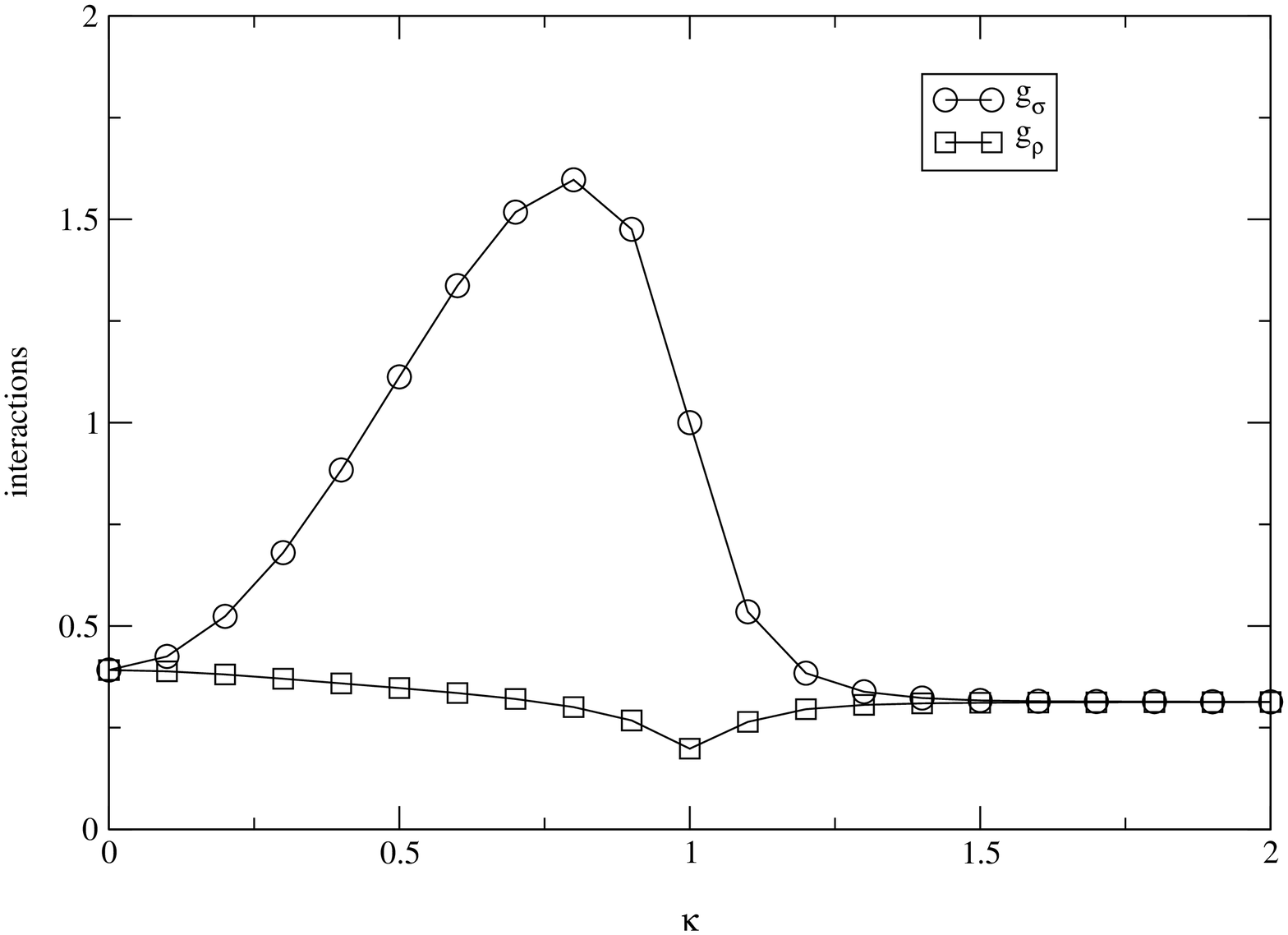}}} \par}

\caption{\label{rgflowhalffilling}RG flow at half-filling as a function of
\protect\( \kappa \protect \).}
\end{figure}

A numerical solution of these sets of equations is shown in Fig.~(\ref{numericalsolutionouthalf})
and Fig.~(\ref{rgflowhalffilling}). For these particular plots we
used \( y_{1,2,3}\left( 0\right) =0.01 \), \( g_{\sigma }\left( 0\right) =g_{\rho }\left( 0\right) =1 \)
and \( \mathrm{G}\left( 0\right) =\tilde{G}\left( 0\right) =\Gamma \left( 0\right) =0 \).
The RG flows were stopped when any of the fugacities reached the value
of \( 1 \) and the values of the other parameters were then plotted
at this point. The flow equations depend only on the absolute value
of \( \kappa  \).

As can be readily checked, the equations always flow to strong coupling.
Nevertheless, special values of \( \left| \kappa \right|  \) allow
us to trace regions with qualitatively different flows. Since the
RG equations depend only on \( \left| \kappa \right|  \), those regions
are mirror reflections on the \( \kappa =0 \) line, the {}``Toulouse
line''.\cite{Zachar-Emery-Kivelson} For \( \kappa ^{2}>3 \), single
spin flip processes are irrelevant (\( y_{1,2}\rightarrow 0 \)),
just like in the FM single impurity Kondo problem.\cite{Yuval-Anderson3,Leggett}
Moreover, the final flow is independent of the precise value of \( \kappa  \)
, clearly indicating a distinct phase of the model. From now on, we
will denote this phase as region \( 1 \). In contrast, spin flips
are always relevant for \( \left| \kappa \right| <3 \), but we also
encounter a second special flow. For \( \left| \kappa \right| =1 \),
the particle fugacities \( G \) and \( \tilde{G} \) are always the
same. There is also a precise balance between the {}``magnetic''
(\( \kappa ^{2}/g_{\sigma } \)) and {}``electric'' (\( g_{\sigma } \))
interactions. Consequently, the ground state is a plasma for particles
of type \( \left( m,e,0\right)  \), implying that \( \phi _{s} \)
and \( \theta _{s} \) are completely disordered. In fact, \( \kappa =1 \)
corresponds to the critical point of the problem of two weakly coupled
LL's. Therefore, we can safely identify \( \left| \kappa \right| =1 \)
as a boundary between different phases. For other values of \( \left| \kappa \right|  \)
the interactions are screened (\( g\rightarrow 0\, \textrm{or}\, \infty  \))
and/or the fugacities have different flows. It is clear that for \( 1<\kappa ^{2}<3 \)
(denoted as region \( 2 \)) single-spin-flip fugacities become less
and less relevant as \( \kappa ^{2}\to 3 \). This suggests a transition
region from the disordered state at \( \left| \kappa \right| =1 \)
to the flow of region \( 1 \). We shall call \( \left| \kappa \right| <1 \)
region \( 3 \). In contrast to the previous cases, single flips are
always strongly relevant in this region. The order of relevance of
the fugacities changes a few times as \( \kappa  \) is varied in
this region. However, a particularly simple case occurs in the {}``Toulouse
line'' (\( \kappa =0 \)).

Even though the renormalization flows are clear and the special flows
were identified, their physical interpretation is less straightforward.
In order to proceed we must assign a physical meaning to each particle
in the gas, from which we can then attempt to determine the phase
diagram.

\section{Effective Hamiltonians\label{eff-ham}}

At each RG step we rewrote the problem as a CG. Moreover, all the
neutrality conditions were preserved by the RG step. We therefore
can define a quantum Hamiltonian that reproduces the CG at each step.
This effective Hamiltonian allows us to understand the behavior of
the system and, in certain special cases, to infer its phase.

In the dense limit of the KLM, the distance between localized spins
is of the order of the smallest bosonic wavelength available. Therefore,
after the first RG step we were forced to introduce new entities in
the problem. Their Hamiltonian form is trivially guessed from their
definitions, \begin{subequations}\label{fusionop}\begin{eqnarray}
O_{1} & \sim  & \tilde{G}\left[ F^{\dagger }_{R\uparrow }F^{\phantom {\dagger }}_{R\downarrow }F^{\dagger }_{L\uparrow }F^{\phantom {\dagger }}_{L\downarrow }\right. \nonumber \\
 &  & \left. e^{i\left| \kappa \right| \sqrt{\frac{8\pi }{g_{\sigma }}}\theta _{s}\left( x\right) }S^{-}\left( x+\delta \right) S^{-}\left( x\right) +h.c.\right] ,\label{o1} \\
O_{2} & \sim  & G\left[ F^{\dagger }_{R\uparrow }F^{\phantom {\dagger }}_{R\downarrow }F^{\dagger }_{L\downarrow }F^{\phantom {\dagger }}_{L\uparrow }\right. \nonumber \\
 &  & \left. e^{i\sqrt{8\pi g_{\sigma }}\phi _{s}\left( x\right) }S^{+}\left( x+\delta \right) S^{-}\left( x\right) +h.c.\right] ,\label{o2} \\
O_{3} & \sim  & 4\Gamma \sum _{\left\{ \eta \neq \nu \right\} =R,L}\sum _{\left\{ \sigma _{1},\sigma _{2}\right\} =\uparrow ,\downarrow }\left[ F^{\dagger }_{\eta \sigma _{1}}F^{\phantom {\dagger }}_{\nu \sigma _{1}}F^{\dagger }_{\nu \sigma _{2}}F^{\phantom {\dagger }}_{\eta \sigma _{2}}\right. \nonumber \\
 &  & \left. e^{i\sqrt{8\pi g_{\rho }}\phi _{c}\left( x\right) +4ik_{F}x}S^{z}\left( x+\delta \right) S^{z}\left( x\right) +h.c.\right] ,\label{o3} 
\end{eqnarray}
\end{subequations}where \( \delta  \) is a distance of the order
of the inverse of the bosonic cut-off \( \left( \sim \alpha \right)  \).
Both the \( O_{1} \) and the \( O_{2} \) terms involve simultaneous
flips of two nearby spins and the creation of particles with charges
\( (m,e,c)=(\pm 2,0,0) \) and \( (0,\pm 2,0) \), respectively. In
contrast, \( O_{3} \) is not related to spin flips and generates
particles with charges \( \left( 0,0,\pm 2\right)  \). It is simple
to understand their origin. In the original Hamiltonian of Eq.~(\ref{bosonham}),
it is possible to spatially resolve the fermion-spin scattering events
. As we reduce the bosonic cut-off this is no longer true, and we
must consider multiple scattering events within the new smallest scale,
\( \alpha  \). There are clear similarities between the conduction
electron operators in Eqs.~(\ref{fusionop}) and the usual backscattering
and Umklapp operators. The standard picture of the RKKY interaction
is that of an effective spin-spin interaction mediated by the conduction
electrons. In light of Eqs.~(\ref{fusionop}), it seems natural to
consider also the opposite point of view: an indirect electron-electron
interaction mediated by the local spins. The RG procedure introduces
these composite events in a natural fashion.

The final operator that must be introduced in the effective Hamiltonian
is a result of the annihilation process. Unlike fusion, when a pair
is annihilated the zeroth order term in an Operator Product Expansion
of the bosonic fields is a constant. Nevertheless, it is still a function
of the local spins and must be considered at the last RG step in order
to establish an effective Hamiltonian. Collecting all possible pair
annihilation terms and expanding point-split bosonic operators we
get

\begin{eqnarray}
O_{z} & \sim  & 4\left( \tilde{G}^{2}-G^{2}\right) S^{z}\left( x+\delta \right) S^{z}\left( x\right) \nonumber \\
 & + & \left( y_{1}^{2}+y_{2}^{2}\right) S^{-}\left( x+\delta \right) S^{+}\left( x\right) +\mathrm{h}.\, \mathrm{c}.\label{annihop} 
\end{eqnarray}

\begin{figure}
{\centering \resizebox*{3in}{!}{\includegraphics{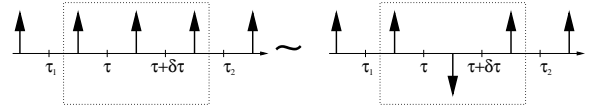}} \par}

\caption{\label{smallestdistance}Two different spin histories at the older
RG scale that cannot be distinguished at the new one.}
\end{figure}

It must be stressed that the spin operators in Eqs.~(\ref{fusionop})
and (\ref{annihop}) should not be understood as the original local
spins. Consider the spin history of Fig~\ref{smallestdistance} as
an example. Suppose that the pair flip-antiflip is produced by a forward
\( J_{\perp }^{f} \) term and a backward \( J_{\perp }^{b} \) term
at times \( \tau  \) and \( \tau +\delta  \) within the new renormalization
scale. This is equivalent to having no flip at all and cannot be distinguished
from a particle with fugacity \( J_{z}^{b} \). At the operator level,
this is formally accomplished by summing over all possible products
of flip operators, expanding the result in \( \delta \sim \alpha  \)
and reordering the Klein factors. The latter are actually crucial
for the correct final sign (see Appendix~\ref{RGexample} for details).
Thus, we exactly reproduce the \( z \)-backscattering particle by
\emph{defining the \( S^{z} \) spin at the new scale} as

\[
2S^{z}\left( \bar{x},\bar{\tau }\right) \equiv S^{+}\left( x,\tau +\delta \right) S^{-}\left( x,\tau \right) -S^{-}\left( x,\tau +\delta \right) S^{+}\left( x,\tau \right) .\]
A similar calculation can be done for any other possible spin history
and bosonic operator within a disk of radius \( \delta \sim \alpha  \).
Therefore, the local spins in the effective Hamiltonian represent
block spins (as in the example above) and not the original ones.\\

Taking \( \delta  \) as the lattice spacing at the last RG step and
collecting all these operators, we find the effective Hamiltonian

\begin{widetext}

\begin{eqnarray}
H_{eff} & = & H_{0}+\sum _{j}\left\{ 4\left[ \tilde{G}^{2}-G^{2}\right] +8\Gamma \cos \left( \sqrt{8\pi g_{\rho }}\phi _{c}\left( x_{j}\right) +4k_{F}x_{j}\right) \right\} S^{z}\left( x_{j+1}\right) S^{z}\left( x_{j}\right) \nonumber \\
 & + & 8y_{3}\sin \left( \sqrt{2\pi g_{\rho }}\phi _{c}\left( x_{j}\right) +2k_{F}x_{j}\right) \sin \left( \sqrt{2\pi g_{\sigma }}\phi _{s}\left( x_{j}\right) \right) S^{z}\left( x_{j}\right) \nonumber \\
 & + & 2\left[ y_{1}\cos \left( \sqrt{2\pi g_{\sigma }}\phi _{s}\left( x_{j}\right) \right) +y_{2}\cos \left( \sqrt{2\pi g_{\rho }}\phi _{c}\left( x_{j}\right) +2k_{F}x_{j}\right) \right] e^{i\sqrt{\frac{2\pi }{g_{\sigma }}}\kappa \theta _{s}\left( x_{j}\right) }S^{+}\left( x_{j}\right) \nonumber \\
 & + & \left[ Ge^{i\sqrt{8\pi g_{\sigma }}\phi _{s}\left( x_{j}\right) }+y_{1}^{2}+y_{2}^{2}\right] S^{+}\left( x_{j+1}\right) S^{-}\left( x_{j}\right) \nonumber \\
 & + & \tilde{G}e^{i\sqrt{\frac{8\pi }{g_{\sigma }}}\kappa \theta _{s}\left( x_{j}\right) }S^{+}\left( x_{j+1}\right) S^{+}\left( x_{j}\right) +\mathrm{h}.\, \mathrm{c}.\label{effectivehamiltonian} 
\end{eqnarray}

\end{widetext}

\section{1D anisotropic KLM phase diagram\label{1dklsec}}

For certain values of \( \kappa  \) the effective Hamiltonian in
Eq.~(\ref{effectivehamiltonian}) is independent of the bosonic fields
at the end of the RG flow. We will exploit these cases to intuit the
various phases of the model.

We start by considering the system away from half-filling, where \( y_{2,3} \)
are irrelevant and \( \Gamma \equiv 0 \). The RG flows are summarized
in Table~\ref{tabela1}.

\begin{table}
\begin{tabular}{|c|c|c|c|c|}
\hline 
 Region&
 \( y_{1} \)&
 \( G \)&
 \( \tilde{G} \)&
 \( \tilde{G}-G \)\\
\hline
\hline 
1&
 \( \rightarrow 0 \)&
 \( \rightarrow \infty  \)&
 \( 0 \)&
 \( <0 \)\\
\hline
2&
 \( \rightarrow \infty  \)&
 \( \rightarrow \infty  \)&
 \( \rightarrow \infty  \)&
 \( <0 \)\\
\hline
3&
 \( \rightarrow \infty  \)&
 \( \rightarrow \infty  \)&
 \( \rightarrow \infty  \)&
 \( >0 \)\\
\hline
\end{tabular}

\vspace{0.2cm}

\caption{\label{tabela1}RG flows for the fugacities away from half-filling.}
\end{table}

In region 1, the only relevant fugacity is \( G \). Therefore, \( \sqrt{8\pi g_{\sigma }}\phi _{s} \)
freezes at \( \pi  \). This reduces Eq.~(\ref{effectivehamiltonian})
to the anisotropic ferromagnetic Heisenberg model

\begin{eqnarray}
H_{eff} & \sim  & \sum _{j}\left( -4G^{2}\right) S^{z}\left( x_{j+1}\right) S^{z}\left( x_{j}\right) \label{hameffFM} \\
 & - & G\left[ S^{+}\left( x_{j+1}\right) S^{-}\left( x_{j}\right) +S^{+}\left( x_{j}\right) S^{-}\left( x_{j+1}\right) \right] \nonumber 
\end{eqnarray}
in its ordered phase (\( G\sim 1,\left\langle S^{z}\right\rangle =1/2 \)). 

The effective Hamiltonian for the \( \kappa =0 \) line, the {}``Toulouse''
line, is also independent of the bosonic field. Since the most relevant
fugacity is \( y_{1} \), \( \sqrt{2\pi g_{\sigma }}\phi _{s} \)
freezes at \( \pi  \). This leads to an antiferromagnetic XYZ model
in an external field

\begin{eqnarray}
H_{eff} & \sim  & \sum _{j}2\left[ \tilde{G}^{2}-G^{2}\right] S^{z}\left( x_{j+1}\right) S^{z}\left( x_{j}\right) -4y_{1}S^{x}\left( x_{j}\right) \nonumber \\
 & + & \left( \frac{y_{1}^{2}+G+\tilde{G}}{2}\right) S^{x}\left( x_{j+1}\right) S^{x}\left( x_{j}\right) \nonumber \\
 & + & \left( \frac{y_{1}^{2}+G-\tilde{G}}{2}\right) S^{y}\left( x_{j+1}\right) S^{y}\left( x_{j}\right) .\label{xkelineouthalffilling} 
\end{eqnarray}
 In this case, the effective spin Hamiltonian exhibits order in the
XY plane, \( G\sim \widetilde{G}\sim y\sim 1 \). Nevertheless, this
does not imply any order of the original spins. As we stated before,
all our results must be understood in the rotated basis of Eq.~(\ref{ZKErotation}).
This ensures that the original model, Eq.~(\ref{bosonham}), is still
disordered, as emphasized in Ref.~\onlinecite{Zachar-Emery-Kivelson}.
Therefore, the system is paramagnetic with short range antiferromagnetic
correlations. Although the {}``Toulouse line'' corresponds to a
particular case, it seems reasonable to extend this assignment to
the entire region \( 3 \). For one thing, because the first term
in Eq.~(\ref{xkelineouthalffilling}), which drives the antiferromagnetic
tendency, remains positive throughout this region. Besides, the XY
disordering terms are the dominant interaction in the region. In particular,
in the \( \left| \kappa \right| =1 \) line, the symmetric flow of
\( G \) and \( \tilde{G} \) ensures that the \( z \)-term vanishes
and therefore the order parameter \( \left\langle S^{x,y,z}\right\rangle  \)
is still zero. Hence, we propose that the entire region 3 is a paramagnetic
phase with short range antiferromagnetic correlations. Note that this
is not necessarily true for other observables, since the flows near
\( \left| \kappa \right| =0 \) and \( \left| \kappa \right| =1 \)
are qualitatively different. 

There is no simple effective Hamiltonian within region 2, but the
disordering term, proportional to \( y_{1} \), becomes progressively
less relevant as \( \kappa ^{2}\to 3 \). More importantly, the short
range \( z \) correlations turn from antiferro- to ferromagnetic.
Consistent with the identification of region \( 1 \) as a ferromagnetic
phase, these two features lead us to tentatively identify region 2
as a ferromagnetically ordered phase with unsaturated magnetization
of the spins.

\begin{figure}
{\centering \resizebox*{4in}{!}{\includegraphics{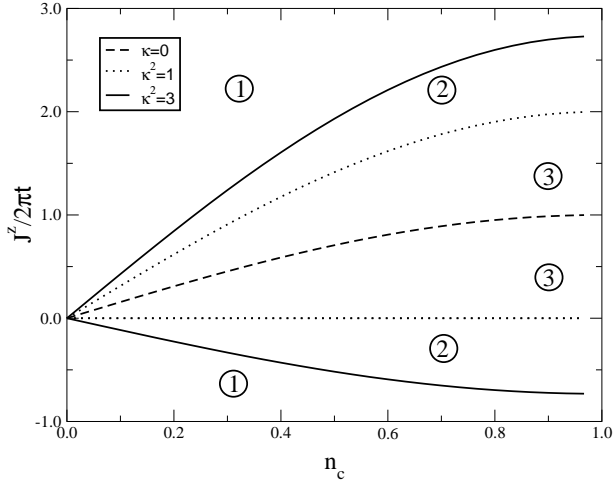}} \par}

\caption{\label{phasediagram}Phase diagram of the 1D anisotropic KLM away
from half-filling. Regions 1 are fully polarized ferromagnets, regions
2 are partially polarized ferromagnets and regions 3 are paramagnetic.
See text for details.}
\end{figure}

Collecting these results, we conclude that there are at least two
continuous phase transitions in the anisotropic KLM far from half-filling.
The first transition, from region 1 to region 2 in Fig.~\ref{phasediagram},
reminiscent of the Berezinskii-Kosterlitz-Thouless transition of the
single impurity Kondo model,\cite{Yuval-Anderson3} separates regions
of relevance and irrelevance of the single flip process. The effective
model for region 1, Eq.~(\ref{hameffFM}), has ferromagnetic order
with full saturation of the \emph{localized} spins. A regime with
ferromagnetic order, however, is beyond the present bosonization treatment,
since the spin polarization of the conduction electrons leads to different
Fermi velocities for up and down spin electrons. However, the RG flow
is still able to indicate its existence through the irrelevance of
single spin flips and the nature of the effective Hamiltonian (\ref{hameffFM}).
Ferromagnetism is simple to understand in the strong coupling limit
(\( \left| J_{z}\right| \gg J_{\perp } \)). In this case, ferromagnetic
ordering allows the electrons to lower their kinetic energy. In fact,
this picture seems to survive down to the isotropic case both for
\( J_{z}>0 \)\cite{Sigrist,Tsunetsugu} and \( J_{z}<0 \).\cite{Dagotto1}
In the FM case, this mechanism is well established and is usually
called double exchange. However, in the AFM chain this simple image
of an up-spin electron moving in a background of localized down-spins
is no longer valid (as can be analytically checked in the Kondo lattice
with one electron\cite{Sigrist,Tsunetsugu}). In the latter, the objects
that lower the kinetic energy are actually the Kondo singlets. Following
this argument, the total spin per site (electrons+spins) would be
\( S_{tot}^{z}=\left\langle S^{z}\right\rangle -n_{c}/2=(1-n_{c})/2 \)
in the antiferromagnetic case and \( S_{tot}^{z}=(1+n_{c})/2 \) in
the ferromagnetic one, as observed numerically.\cite{Tsunetsugu,Dagotto1}

There is another continuous phase transition line from region 2 to
region 3 in Fig.~\ref{phasediagram}, similar to the transition of
the Ising model in a transverse field,\cite{Kogut1} that separates
a paramagnetic phase (region 3 of Fig.~\ref{phasediagram}) from
a region with unsaturated magnetization of the localized spins. The
magnetization grows continuously up to the border of region 1. It
is tempting to identify region 2 with similar phases with unsaturated
moments found in numerical studies of both the isotropic FM KLM of
Dagotto \textit{et al.}\cite{Dagotto1} and the isotropic AFM KLM
of Tsunetsugu \textit{et al.}\cite{Tsunetsugu}

The numerical studies of the FM KLM also identified a region of phase
separation.\cite{Dagotto1} We did not find any indication of phase
separation. We can think of two reasons why. First, the coupling constant
in that region is of the order of the electron bandwidth and therefore
bosonization is no longer valid. Moreover, this phase is a competition
between the ferromagnetic tendencies of lower band fillings and the
antiferromagnetic counterpart at half-filling. Since we completely
neglect backscattering (ultimately responsible for the antiferromagnetism)
this phase was lost even before we began.

As we pointed out before, at half-filling we are able to include the
backscattering terms in the RG scheme. We will now consider this case.
The first result from the RG flow is that \( O_{3} \), whose bosonic
part is identical to an Umklapp term, is always relevant, pointing
to the presence of a charge gap. The spin sector is more subtle and
we must consider some special cases.

Region 1 can be simply analyzed. The interaction parameters \( g_{\sigma } \)
and \( g_{\rho } \) go to zero and the most relevant fugacity is
\( y_{3} \). Therefore, \( \sqrt{2\pi g_{\rho }}\phi _{c} \) freezes
at \( \frac{\pi }{2} \). The other relevant flows are \( \Gamma  \)
and \( G \). The effective Hamiltonian reduces to an anisotropic
ferromagnetic Heisenberg model in a staggered field,

\begin{eqnarray*}
H_{eff} & \sim  & \sum _{j}\left( -4G^{2}-8\Gamma \right) S^{z}\left( x_{j+1}\right) S^{z}\left( x_{j}\right) \\
 & - & GS^{+}\left( x_{j+1}\right) S^{-}\left( x_{j}\right) +\mathrm{h}.\, \mathrm{c}.\\
 & + & \left( -1\right) ^{x_{j}}8y_{3}S^{z}\left( x_{j}\right) .
\end{eqnarray*}
The staggered field induces Néel order, but the system has strong
ferromagnetic tendencies. As we move away from half-filling, the staggered
field becomes progressively irrelevant and the ferromagnetic effective
model is reobtained.

The \( \kappa =0 \) point is once again very special. At the end
of the RG flow, the effective Hamiltonian is also free of the bosonic
fields and the most relevant fugacities are \( y_{2} \) and \( y_{1} \).
They force \( \sqrt{2\pi g_{\rho }}\phi _{c} \) and \( \sqrt{2\pi g_{\sigma }}\phi _{s} \)
to freeze at \( \pi  \), suppressing the staggered field in the \( z \)
direction. Thus, the {}``Toulouse point'' effective Hamiltonian is

\begin{eqnarray*}
H_{eff} & \sim  & \sum _{j}\left[ 2\left( \tilde{G}^{2}-G^{2}\right) +4\Gamma \right] S^{z}\left( x_{j+1}\right) S^{z}\left( x_{j}\right) \\
 & - & \left( \frac{G+y_{1}^{2}+y_{2}^{2}+\tilde{G}}{2}\right) S^{x}\left( x_{j+1}\right) S^{x}\left( x_{j}\right) \\
 & - & \left( \frac{G+y_{1}^{2}+y_{2}^{2}-\tilde{G}}{2}\right) S^{y}\left( x_{j+1}\right) S^{y}\left( x_{j}\right) \\
 & - & 4\left( y_{1}+\left( -1\right) ^{x_{j}}y_{2}\right) S^{x}\left( x_{j}\right) .
\end{eqnarray*}
As before this does not imply any order of the original spins in the
XY plane. From this effective model we can see that the \( \left| \kappa \right| =0 \)
point is characterized by spin and charge gaps and no ordering.

In summary, at half filling we assign two distinct magnetic phases.
Regions \( 1 \) and \( 2 \) have Néel order in the \( z \) direction.
On the other hand, if we assume that the {}``Toulouse line'' features
can be extended to the entire region \( 3 \), we can identify this
region with a paramagnetic phase. The several changes in the relative
flows maybe a sign of additional phases as a function of \( \kappa  \).
However, the effective Hamiltonian cannot be so easily solved and
we are unable to make further progress.

The KLM at half-filling was studied by Shibata \emph{et al.}\cite{Shibata1}
By looking at the strong coupling limit, they were able to find five
distinct magnetic phases, which they argue survive down to weak coupling:
two Néel phases (FM and AFM), a planar phase (the triplet state with
\( S^{z}=0 \)), a Haldane phase and a Kondo singlet (paramagnetic)
phase. Because of the relevance of backscattering and the condition
\( \tilde{J}_{z}^{b}\ll 1 \), a direct comparison between the RG
flows and the available numerical results is restricted to \( \kappa \sim 1 \).
This neighborhood has no simple effective Hamiltonian and we are unable
to make direct contact with the numerical results. We can point out,
however, that the strong coupling flow of \( g_{\sigma } \) is an
indication of the opening of a spin gap, though this is less certain
because of the difficulty of analyzing the effective Hamiltonian.
This possible spin gap is compatible with the Haldane type phase at
\( J_{z}<0 \) and the Kondo singlet phase at \( J_{z}>0 \) obtained
in Ref.~\onlinecite{Shibata1}. As we dope the system away from half-filling
the backward-scattering terms become irrelevant, and a direct comparison
with the numerical data becomes more feasible.

As a final illustration of the usefulness of the Coulomb gas mapping,
we develop in Section~\ref{ising-kondo} its application to a related
yet simplified model of spins and fermions: the Ising-Kondo chain.
Its simplicity makes it a more pedagogical example of the formalism.

\section{The Ising-Kondo chain\label{ising-kondo}}

The Ising-Kondo model, 

\[
H=\sum _{\vec{k},\sigma }\varepsilon _{\vec{k}}\psi ^{\dagger }_{\vec{k},\sigma }\psi _{\vec{k},\sigma }+J\sum _{i,s,\acute{s}}S_{i}^{z}\psi ^{\dagger }_{i,s}\frac{\sigma _{s,\acute{s}}^{z}}{2}\psi _{i,\bar{\sigma }}+y\sum _{i}S_{i}^{x},\]
was proposed by Sikkema \emph{et al.}\cite{Sikkema1} as a model for
the weak antiferromagnetism of \( \mathrm{URu}_{2}\mathrm{Si}_{2} \).
Here, we will consider the one dimensional version of this model and
apply the same methods that we used in the Kondo chain. Using bosonization
and disregarding the backscattering terms, the Hamiltonian simplifies
to

\begin{equation}
\label{hamiltonianajahnteller1}
H=H_{0}+\sum _{i}\sqrt{\frac{2}{\pi }}\tilde{J}\partial _{x}\phi \left( i\right) S^{z}\left( i\right) -\tilde{y}S^{x}\left( i\right) ,
\end{equation}
where the coupling constants were rescaled by the Fermi velocity as
before. Eq.~(\ref{hamiltonianajahnteller1}) is identical to the
co-operative Jahn-Teller Hamiltonian.\cite{Gehring} The lower symmetry
of the model allows us to foresee that the sign of \( J \) is irrelevant
to the physics. It is also a well known result from the co-operative
Jahn-Teller problem that the strong coupling limits \( \tilde{J}\gg 1 \)
and \( \tilde{y}\gg 1 \) show easy axis order in the \( z \) and
\( x \) directions, respectively.

Exactly as in the KLM, we can proceed by going to a path integral
formulation with bosonic coherent states and the local spin \( S_{z} \)
basis. After tracing the bosonic fields and integrating by parts the
spin variables, the Coulomb gas that follows has only one breed of
particles \( \left( m,0,0\right)  \), subjected to the neutrality
condition 1 of Section~\ref{sectionpartitionfunction}. To mimic
our previous notation we define \( g=\sqrt{\frac{2\pi }{J}} \). Assuming
\( \tilde{y}\ll 1 \), the RG equation can be derived in a similar
fashion. They correspond to the standard Kosterlitz-Thouless equations,

\begin{eqnarray*}
\frac{dy}{dl} & = & 2\left( 1-g\right) y,\\
\frac{d\ln g}{dl} & = & -gy^{2}.
\end{eqnarray*}

\begin{figure}
{\centering \resizebox*{4in}{!}{\includegraphics{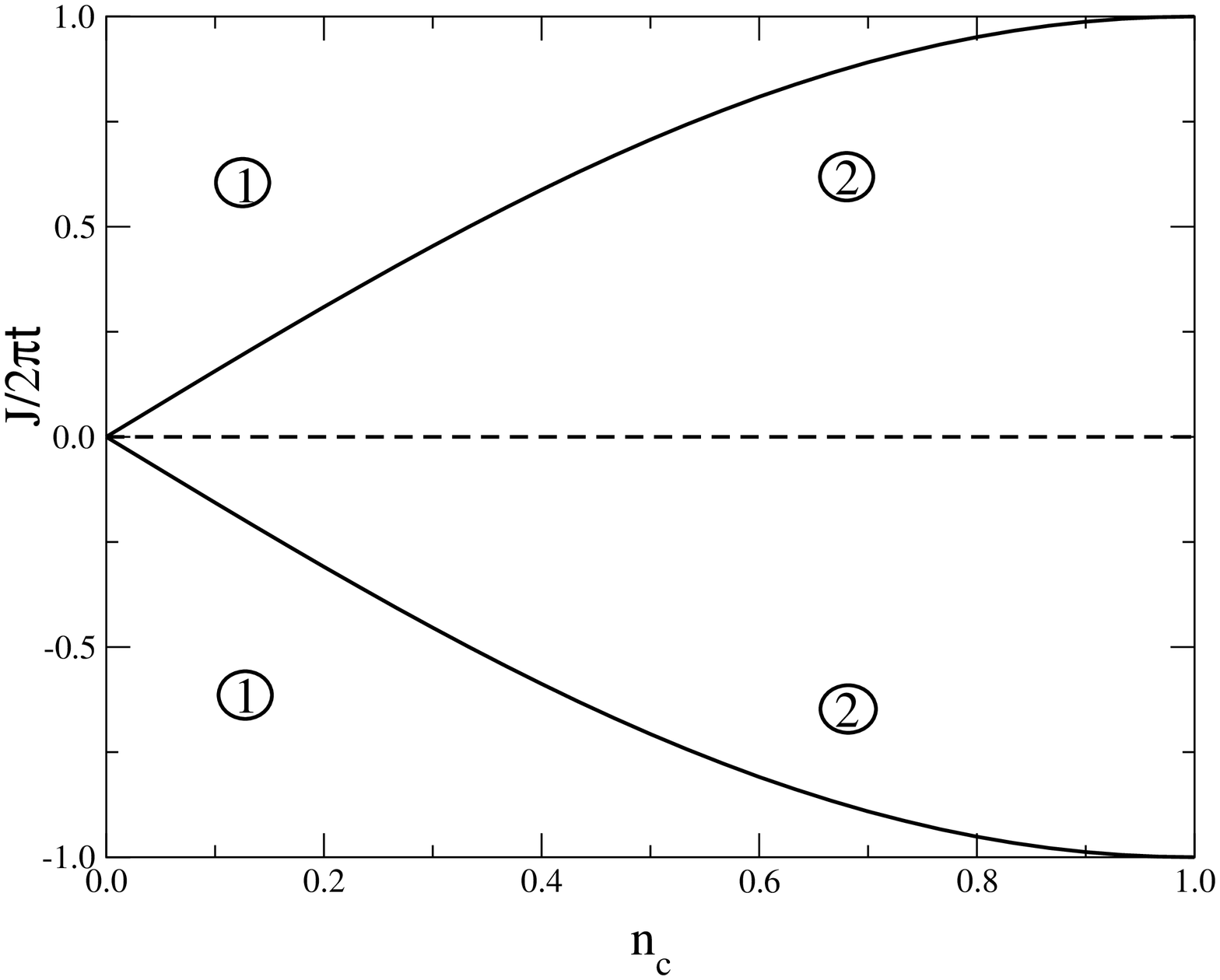}} \par}

\caption{\label{phaseIsingKondo}Phase diagram of the Ising-Kondo
  chain. In Region~1 the transverse field is irrelevant, while in Region~2
  it is relevant. See text for details.}
\end{figure}

For \( g>1 \), spin flip processes are irrelevant (see Fig.~\ref{phaseIsingKondo},
region 1). In the Jahn-Teller language this corresponds to a ferrodistortion
of the \( y\ll J \) fixed point. On the other hand, for \( g<1 \)
spin flips are relevant, \( y\rightarrow \infty  \) and \( g\rightarrow 0 \)
(see Fig.~\ref{phaseIsingKondo}, region 2). We can find an effective
Hamiltonian to shed light on the physics in this regime. Indeed, we
could have applied the rotation in Eq.~(\ref{ZKErotation}) to the
original Hamiltonian to get\[
H=H_{0}-\frac{y}{2}e^{-i\sqrt{8\pi g}\theta \left( x_{i}\right) }S^{+}\left( x_{i}\right) +\mathrm{h}.\, \mathrm{c}.+\mathrm{s}.\, \mathrm{r}.\, \mathrm{t}.,\]
 where {}``s. r. t.'' stands for {}``short range terms''. The
operators in this rotated basis are called {}``displaced'' in the
co-operative Jahn-Teller literature.\cite{Gehring} This rotation
is equivalent to the integration by parts of the \( S_{z} \) variables
in the time direction, as we saw. By taking now \( y\rightarrow \infty  \)
and \( g\rightarrow 0 \), the effective Hamiltonian is simply a magnetic
field in the \( x \) direction acting to order the local spins. Unlike
in the KLM, the original spins are also ordered in the \( x \) direction
since \( \theta  \) freezes at the value of zero. The transition
is continuous and of the Kosterlitz-Thouless type.

\section{Discussion and Conclusions\label{discconc}}

We have proposed in this article what is the natural extension to
the one-dimensional lattice of the highly successful approach of Anderson,
Yuval and Hamman\cite{Yuval-Anderson1,Yuval-Anderson2,Yuval-Anderson3}
to the single impurity Kondo problem. The mapping to a Coulomb gas
is made specially easy by using bosonization methods and particularly
subtle developments demonstrate the importance of a careful consideration
of Klein factors, so often neglected in most treatments.\cite{Delft-Schoeller}
Since bosonization relies on the linearization of the conduction electron
dispersion and is appropriate for the analysis of the long-wavelength
physics it is never quite obvious how far it can be taken in its application
to lattice systems. However, motivated by its success in the Hubbard,
Heisenberg and other models, it is reasonable to attempt a direct
comparison of our treatment to the phases of the anisotropic Kondo
lattice model.

One of the hardest tasks in our treatment is the extraction of physical
information from the effective models we obtain after several rescaling
steps. Some special lines in the phase diagram can be more confidently
analyzed, but as is common in RG treatments, we are then forced to
attempt an extrapolation to other regions based on continuity arguments.
This is specially true in our case, where most of the flows are towards
strong coupling. Given these caveats, however, the overall topology
of the phase diagram away from half-filling is compatible with the
known phases of the \emph{isotropic} model.\cite{Tsunetsugu,Dagotto1}
The extension of these studies to the anisotropic case would be highly
desirable. At half-filling, the method itself limits its application
to the \( J_{z}\ll t \) region. Unfortunately, this is one of the
regions where the effective Hamiltonian is hard to solve and we are
not able to explore the rich phase diagram obtained in Ref.~\onlinecite{Shibata1}.
Nevertheless, we do find a charge gap at half-filling throughout the
phase diagram, which seems compatible with the numerical results.
The question of the spin gap is less clear but our results are also
compatible with what is known numerically.

We would also like to try to make contact with previous studies of
the Kondo lattice model in one dimension based on the use of Abelian
bosonization. In the important work of Zachar, Kivelson and Emery,\cite{Zachar-Emery-Kivelson}
where the rotation of Eq.~(\ref{ZKErotation}) is first used, the
highly anisotropic {}``Toulouse line'' (\( \kappa =0 \)) is analyzed
in detail. One of their findings is the presence of a spin gap in
the spectrum away from half-filling, which also appears in our effective
Hamiltonian. At half-filling, they also find spin and charge gaps,
which seem compatible with our results. They also point out that the
metal-insulator transition as \( n\to 1 \) is of the commensurate-incommensurate
type.\cite{schulz4} In our treatment, the commensurability condition
\( 4k_{F}a=2\pi n=1 \) for the relevance of the backward scattering
terms, the same as in the Hubbard model, is a strong indication that
the transition is indeed of this type.

Honner and Gulácsi have also investigated the spin dynamics of the
isotropic Kondo chain.\cite{Gulacsi1,Gulacsi2} After smearing out
the discontinuity in the commutation relations of the bosonic fields,
they replace the latter by their expectation value in the non-interacting
ground state and write an effective Hamiltonian for the localized
spins. This Hamiltonian can then be treated numerically and the phase
diagram determined. This procedure requires the fitting of the smearing
length scale to numerical results. One of the advantages our treatment
brings to the problem is the ability to do the full analysis analytically
and without any \emph{a priori} assumption about the boson dynamics.
In fact, the Coulomb gas mapping treats spins and bosons on the same
footing. Besides, no fitting to numerical results is necessary. A
discrepancy between our results and those of Honner and Gulácsi is
the partially polarized FM phase we find at \( J_{z}<0 \). In their
treatment, a PM phase is found instead. It would be interesting to
extend their treatment to the anisotropic case for a fuller comparison.

Recently, Zachar\cite{Zac01} conceived an alternative approach to
the KLM in the rotated basis. He used a particular example of the
rotation in Eq.~(\ref{ZKErotation})\[
\bar{U}=e^{i\sqrt{2\pi }\sum _{x}\theta _{s}\left( x\right) S^{z}\left( x\right) },\]
and treated the KLM in a self-consistent mean-field approximation.
This approach led him to predict three different phases in the AFM
KLM as well. The first region is controlled by the paramagnetic {}``Toulouse
line'' fixed point. In the rotated basis, this phase is characterized
by \( \left\langle S^{z}\left( x\right) \right\rangle =0 \) and \( \left\langle S^{x}\neq 0\right\rangle  \),
precisely as we find in region \( 3 \) of Fig.~\ref{phasediagram}.
Another phase has \( \left\langle S^{z}\right\rangle \neq 0 \) and
\( \left\langle S^{x}\right\rangle =0 \). In this case, the system
exhibits ferromagnetic order in the original basis, and therefore
could be identified with region \( 1 \) of Fig.~\ref{phasediagram}.
Finally, embedded between these two phases, he also finds a third
intermediate region, which he identifies as a {}``soliton lattice'',
with \( \left\langle S^{z}\right\rangle \neq 0 \) and \( \left\langle S^{x}\right\rangle \neq 0 \).
It is tempting to associate this intermediate phase with region 2
of Fig.~\ref{phasediagram}. However, Zachar proposes a different
description calling region 1 a {}``staggered liquid Luttinger liquid'',
whereas we find it much more natural to associate \( \left\langle S^{z}\right\rangle \neq 0 \)
with ferromagnetic order. He also conjectures that region \( 2 \)
does not exist. Finally, he argued that all transitions are first
order and of the commensurate-incommensurate type, while we find them
to be continuous.

In conclusion, we have presented a flexible treatment of a one-dimensional
system of spins and fermions based on a mapping to a Coulomb gas,
which we treat within a renormalization group approach. When applied
to the Kondo lattice model, the method enables us to identify its
various phases both at and away from half-filling.

\begin{acknowledgments}
We thank I.~Affleck, A.~L.~Chernyshev, E.~Dagotto, M.~Gul\'{a}csi,
N.~Hasselmann, S.~Kivelson, S.~Sachdev, J.~C.~Xavier and O.~Zachar,
for suggestions and discussions. E.~N., E.~M. and G.~G.~Cabrera
acknowledge financial support from FAPESP (01/00719-8, 01/07777-3)
and CNPq (301222/97-5). A.~H.~C.~N. acknowledges partial support
provided by a CULAR grant under the auspices of the US DOE.
\end{acknowledgments}
\appendix

\section{Neutrality Conditions for the KLM\label{neutralityapendix}}

Neutrality conditions are common in Coulomb gas formulations of quantum
problems. In the simplest applications, these conditions impose that
the overall charge is zero as, for instance, in the sine-Gordon model.\cite{Gogolin, Niehnus}
As applied to our case this condition reads

\begin{equation}
\label{neutral1}
\sum _{i}m\left( x_{i},\tau _{i}\right) =\sum _{i}e\left( x_{i},\tau _{i}\right) =\sum _{i}c\left( x_{i},\tau _{i}\right) =0.
\end{equation}
 They are the mathematical expression of the condition for the bosonic
correlation functions \emph{not} to vanish in the thermodynamic limit
and they also ensure the overall cancellation of the Klein factors.
However, in the KLM the presence of both spins and bosons leads to
more stringent neutrality conditions than in other problems. Therefore,
besides Eq.~(\ref{neutral1}), there are two additional restrictions. 

The first one comes from the impossibility of performing two consecutive
upward spin flips on a given localized spin-\( \frac{1}{2} \) site.
Since, from Eq.~(\ref{kinkterms}) the \( m=\pm 1 \) variable gives
the direction of a spin flip, it follows that \( m \) must alternate
in time. This condition is also present in the Coulomb gas formulation
of the single impurity Kondo problem.\cite{Yuval-Anderson1} As a
direct consequence of the alternation of the charge \( m \) and the
periodic boundary conditions in imaginary time, we obtain the first
{}``strong'' neutrality condition: the total charge \( m \) at
a given spatial position is zero \( \sum _{i}m\left( x_{fixed},\tau _{i}\right) =0 \).
This gives condition 1 of Section \ref{sectionpartitionfunction},
whereas condition 2 is already contained in Eq.~(\ref{neutral1}).

The second additional restriction is slightly less obvious. From Eq.~(\ref{partitionfunctionwithbosons}),
we see that each contribution to the partition function has a prefactor
sign that depends on a string of Klein factors and \( S^{z} \) operators,
the latter coming from \( z \) backward-scattering events generated
by \( H^{b}_{z} \) of Eq.~(\ref{backward-z}). The neutrality condition
we will derive comes from the cancellation of terms with identical
absolute values but with opposite prefactor signs. This will finally
lead to condition 3 of Section \ref{sectionpartitionfunction}. We
will now consider different cases separately.

Let us first focus on the contributions to the partition function
coming from terms with forward scattering only (Eqs.~(\ref{forward-z})
and (\ref{forward-perp})). If there are no spin flips, then the prefactor
is obviously positive. When there is a pair of opposite flips (the
only possibility allowed by the neutrality condition 1), then, because
of the overall neutrality condition 2, the Klein factors cancel

\[
F_{\eta \sigma _{1}}^{\dagger }F_{\eta \sigma _{2}}F_{\eta \sigma _{2}}^{\dagger }F_{\eta \sigma _{1}}=1,\]
preserving the positive sign. Consideration of configurations with
additional pairs of flips leads to the same cancellation of Klein
factors.

Next, we look at contributions generated by \( H^{b}_{\perp } \)(Eq.~(\ref{backward-perp}))
only. By considering again increasing numbers of pairs of opposite
flips as in the previous paragraph we arrive at an analogous cancellation
of Klein factors.

Moving on now to contributions coming from \( H_{z}^{b} \) (Eq.~(\ref{backward-z})),
we first consider the possibility of no spin flips. In this case,
integrating out the bosonic modes, the contribution to the partition
function is

\begin{equation}
\label{szszpart}
z\sim \frac{e^{\pm 2ik_{F}\Delta x_{ij}}}{r^{g_{\rho }+g_{\sigma }}_{ij}}S^{z}\left( i\right) S^{z}\left( j\right) ,
\end{equation}
where the Klein factors also cancel nicely. Tracing over the spin
variables leads to no contribution to the partition function sum,
unless \( i \) and \( j \) have the same space coordinate. What
happens for a higher number of insertions of \( H_{z}^{b} \) ? For
the general case of \( N \) particles coming from \( H_{z}^{b} \),
the contribution to the partition function will be\[
z\sim e^{\sum _{ij}\ln \left| r_{ij}\right| \left( g_{\rho }c_{i}c_{j}+g_{\sigma }e_{i}e_{j}\right) }\prod _{i}e^{2ik_{F}c_{i}x_{i}}S^{z}\left( i\right) .\]
 Note how each \( S_{z} \) insertion comes with a corresponding \( c \)
charge. Thus, it is simple to show that tracing over \( S_{z}\left( i\right)  \)
leads to the condition of having \emph{an even number of particles
of charge \( c \) at each spatial coordinate.} Moreover, the reordering
of Klein factors leads to their complete cancellation.

\begin{figure}
{\centering \resizebox*{3in}{!}{\includegraphics{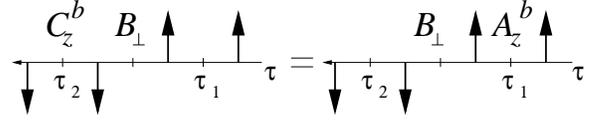}} \par}

\caption{\label{domainwall}A possible spin history and the diagrammatic representation
of Eq.~(\ref{kleinfacID}). In the figure, \protect\( A_{z}^{b}B_{\perp }=B_{\perp }C_{z}^{b}\protect \),
where \protect\( A^{b}_{z}=F_{\eta \sigma }^{\dagger }F_{\nu \sigma }\sigma \left( \tau _{1}\right) \protect \),
\protect\( B_{\perp }=F_{\xi \uparrow }^{\dagger }F_{\psi \downarrow }\protect \)
and \protect\( C^{b}_{z}=F_{\eta \sigma }^{\dagger }F_{\nu \sigma }\sigma \left( \tau _{2}\right) \protect \).}
\end{figure}

In order to generalize this result to a configuration with an arbitrary
number of spin flips let us assume initially that there are only flips
of one kind: either \( H^{f}_{\perp } \) or \( H^{b}_{\perp } \).
By using the identity (using Pauli matrices instead of spin operators)\begin{equation}
\label{kleinfacID}
\sigma _{1}F_{\eta \sigma }^{\dagger }F_{\nu \sigma }F_{\xi \sigma _{1}}^{\dagger }F_{\psi \sigma _{2}}=\sigma _{2}F_{\xi \sigma _{1}}^{\dagger }F_{\psi \sigma _{2}}F_{\eta \sigma }^{\dagger }F_{\nu \sigma },
\end{equation}
with \( \nu \neq \eta  \) and \( \sigma _{1}\neq \sigma _{2} \),
it is easy show that an \( H_{z}^{b} \) insertion on one side of
a domain wall \emph{can be moved to the other side with a sign change}
(see Fig.~\ref{domainwall}). Now consider, for example, a pair of
particles generated by \( H_{z}^{b} \) as before. When there is a
pair of flips lying along the time line, we can move the Klein factors
and \( S^{z} \) through the domain walls with the identity above
and cancel them out. Therefore, our previous result, obtained without
the flips, remains valid. This can be generalized for any number of
flips. Finally, we must consider the possibility of having flips coming
both from \( H_{\perp }^{f} \) and \( H_{\perp }^{b} \). For this
we note that a flip from \( H^{f}_{\perp } \) and a subsequent opposite
flip from \( H^{b}_{\perp } \) {}``fuse'' in a way which is precisely
equivalent to the insertion of a single \( H^{b}_{z} \) particle
(Klein factors, \( S_{z} \) operators and all). Therefore our previous
conclusion is valid in this case as well: there must be an even number
of \( c \) charges (not necessarily neutral) at each space coordinate.

Finally, we have so far considered insertions along one imaginary
time line only, which is not the general case. Nevertheless, because
there is always an even number of Klein factors in each time line,
we can always reorder them so as to group together contributions from
individual time lines without introducing additional signs. Then,
the previous analysis can be used to prove the global cancellation
of Klein factors and \( S_{z} \) operators in the general case as
well.

We would like to note that the arguments presented in this Appendix
indicate a rather surprising precise cancellation of Klein factors
and \( S_{z} \) operators, suggesting that perhaps there is a deeper
underlying symmetry behind this result. However, we were not able
to find a more general symmetry-based demonstration. We also point
out that, in the problem of a single Kondo impurity in a Luttinger
liquid, Lee and Toner\cite{Lee} introduce the same kinds of particles
defined in the Table~\ref{initialparticlestable}. However, in their
analysis there is no explicit mention of how to deal with the product
of Klein factors and the \( S^{z} \) operators coming from the \( z \)
backscattering events. We have shown that these factors almost miraculously
cancel out and do not affect the remainder of the analysis of their
(or our) Coulomb gas.

\section{Annihilation and Fusion of Particles\label{appendixannihilationandfusion}}

We now show in detail how the RG procedure leads to the annihilation
and fusion of charged particles. Consider that we initially have a
{}``close pair'' with each particle having fugacities \( F_{1} \)
and \( F_{2} \). In the complex notation of Eq.~(\ref{zjk}), the
action takes the simple form:

\[
S_{eff}=\frac{1}{2}\sum _{i\neq j}\alpha _{ij}\ln z_{ij}+\beta _{ij}\ln \bar{z}_{ij}+2ik_{F}\sum _{i}c\left( \eta _{i}\right) x_{i},\]
with:

\begin{eqnarray*}
\alpha _{ij} & = & \frac{1}{2}\left[ \left( \frac{\left| \kappa \right| }{\sqrt{g_{\sigma }}}m\left( \eta _{i}\right) -\sqrt{g_{\sigma }}e\left( \eta _{i}\right) \right) \right. \\
 &  & \left. \left( \frac{\left| \kappa \right| }{\sqrt{g_{\sigma }}}m\left( \eta _{j}\right) -\sqrt{g_{\sigma }}e\left( \eta _{j}\right) \right) +g_{\rho }c\left( \eta _{i}\right) c\left( \eta _{j}\right) \right] ,\\
\beta _{ij} & = & \frac{1}{2}\left[ \left( \frac{\left| \kappa \right| }{\sqrt{g_{\sigma }}}m\left( \eta _{i}\right) +\sqrt{g_{\sigma }}e\left( \eta _{i}\right) \right) \right. \\
 &  & \left. \left( \sqrt{g_{\sigma }}e\left( \eta _{j}\right) +\frac{\left| \kappa \right| }{\sqrt{g_{\sigma }}}m\left( \eta _{j}\right) \right) +g_{\rho }c\left( \eta _{i}\right) c\left( \eta _{j}\right) \right] ,
\end{eqnarray*}
 Suppose the {}``close pair'' particles are at positions \( l \)
and \( m \) in space-time. We split the action in three parts:

\begin{eqnarray*}
S_{1} & = & \frac{1}{2}\left( \sum _{i\neq j\neq l}+\sum _{i\neq j\neq m}\right) \left[ \alpha _{ij}\ln z_{ij}+\beta _{ij}\ln \bar{z}_{ij}\right. \\
 & + & \left. 2ik_{F}\sum _{i}c\left( \eta _{i}\right) x_{i}\right] ,\\
S_{2} & = & \sum _{i\neq l}\alpha _{il}\ln z_{il}+\beta _{il}\ln \bar{z}_{il}\\
 & + & \sum _{i\neq m}\alpha _{im}\ln z_{im}+\beta _{im}\ln \bar{z}_{im},\\
S_{3} & = & \alpha _{lm}\ln z_{lm}+\beta _{lm}\ln \bar{z}_{lm}.
\end{eqnarray*}
\( S_{1} \) gives the interaction between the particles which are
not in the {}``close pair'', \( S_{2} \) the interactions between
the {}``close pair'' and the other particles, and \( S_{3} \) the
interaction between the particles belonging to the pair. Finally,
we define the relative coordinate of the pair as: \( s=z_{m}-z_{l} \).
For \( s\sim d\ell \ll 1 \) we expand the logarithm in \( s \):

\[
\ln z_{mi}\cong \ln z_{li}+\frac{s}{z_{li}}.\]

\subsection{Fusion}

If the pair is not neutral, the leading term in the expansion of \( S_{2} \)
is of order zero in \( \left| s\right|  \). Therefore, we can rewrite
\( S_{2} \) as giving the interactions between all other particles
and the new {}``fused'' one. In order to once again write the problem
in a Coulomb gas form, we must rescale the fugacities to accommodate
this new particle. Doing the integral in \( S_{3} \)

\[
\int dse^{S_{3}}\sim \frac{\sin \pi b_{lm}}{b_{lm}}d\ell ,\]
where:

\[
b_{lm}=\kappa \left( e\left( \eta _{m}\right) m\left( \eta _{l}\right) +m\left( \eta _{m}\right) e\left( \eta _{l}\right) \right) .\]
 After summing over particle configurations that do not contain the
fused pair, we get the contribution from fusion of particles with
fugacities \( F_{1} \) and \( F_{2} \) to the fugacity \( F_{3} \)
of this new fused particle

\[
\frac{dF_{3}}{d\ell }=\frac{\sin \pi b_{lm}}{b_{lm}}F_{1}F_{2}.\]

\subsection{Annihilation}

If the pair is neutral, the particles annihilate each other. In this
case, \( \alpha _{il}=-\alpha _{im} \) and \( \beta _{il}=-\beta _{im} \)
. We can expand the partition function contribution in \( \left| s\right|  \)

\[
z=\int dsdl\, e^{S_{1}+S_{3}}\left( 1+\left| s\right| B\left( l,m\right) +\frac{\left| s\right| ^{2}}{2}B\left( l,m\right) ^{2}+...\right) ,\]
 where:

\[
B\left( l,m\right) =\sum _{i\neq \left( l,m\right) }\left( \frac{s}{\left| s\right| }\frac{\alpha _{il}}{z_{il}}-\frac{\bar{s}}{\left| s\right| }\frac{\beta _{il}}{\bar{z}_{il}}\right) .\]
The integration over the pair {}``center of mass'' coordinate \( l \)
is

\[
\int dlB\left( l,m\right) =0.\]
This is different from the single impurity Kondo problem or the dilute
limit, where this integral does not vanish. The reason is that, in
these cases, the integration is only along the time direction and,
therefore, a logarithmic divergence appears. Consequently, the expansion
in \( B \) stops at first order. In contrast, in the dense limit
the integral is over space and imaginary time, hence removing this
singularity. The first non-vanishing term is second order in \( B \),
as in the sine-Gordon and the 2 LL's problem\cite{Gogolin}

\begin{eqnarray*}
\frac{1}{2}\int dlB\left( l,m\right) ^{2} & = & \frac{1}{2}\sum _{i,j\neq \left( l,m\right) }\int dl\, \left[ \frac{s^{2}}{\left| s\right| ^{2}}\frac{\alpha _{il}\alpha _{jl}}{z_{il}z_{jl}}\right. \\
 & + & \left. \frac{\bar{s}^{2}}{\left| s\right| ^{2}}\frac{\beta _{il}\beta _{jl}}{\bar{z}_{il}\bar{z}_{jl}}-2\frac{\alpha _{il}\beta _{jl}}{z_{il}\bar{z}_{jl}}\right] .
\end{eqnarray*}
 After integration, the first two terms are power law functions of
the distance between the remaining particles of the gas. For a sufficiently
dilute gas, the most significant contribution is given by the last
term

\begin{equation}
\label{integralovercentralofmass}
\frac{1}{2}\int dlB\left( l,m\right) ^{2}\sim -2\pi \sum _{i,j\neq \left( l,m\right) }\alpha _{il}\beta _{jl}\ln \left| z_{ij}\right| +\mathrm{const}.
\end{equation}
It has a simple physical meaning: it gives the {}``vacuum polarization''
coming from the dipole moment of the {}``close pair''. The final
step in the calculation is to integrate over the relative coordinate
\( s \)

\begin{equation}
\label{relativecoordcontribuiton}
\int dse^{S_{3}}\left| s\right| ^{2}\sim \frac{\sin \pi b_{lm}}{b_{lm}}d\ell ,
\end{equation}
where

\[
b_{lm}=\kappa \left( e\left( \eta _{l}\right) m\left( \eta _{m}\right) +m\left( \eta _{m}\right) e\left( \eta _{l}\right) \right) .\]
Collecting Eq.~(\ref{integralovercentralofmass}) and Eq.~(\ref{relativecoordcontribuiton}),
the partition function contribution after rescaling is

\[
z=e^{S_{1}}\left( -2\pi \frac{\sin \pi b_{lm}}{b_{lm}}\sum _{i,j\neq \left( l,m\right) }\alpha _{il}\beta _{jl}\ln \left| z_{ij}\right| d\ell \right) .\]
To complete the RG step, we must sum the charge configuration of \( S_{1} \)that
did not contain the {}``close par''

\[
z=e^{S_{1}}\left( 1-2\pi F_{1}F_{2}\frac{\sin \pi b_{lm}}{b_{lm}}\sum _{i,j\neq \left( l,m\right) }\alpha _{il}\beta _{jl}\ln \left| z_{ij}\right| d\ell \right) .\]
Summing over all possible annihilations of pairs of particles and
reexponentiating, we get the renormalization group equations for the
Coulomb interaction strengths \( g_{\sigma } \) and \( g_{\rho } \).

\section{Detailed example of a fusion process\label{RGexample}}

In this Appendix we show in more detail how to interpret the local
spins in the effective Hamiltonian of Section~\ref{eff-ham} after
several RG steps.

Let us focus on the spin histories of Fig.~\ref{smallestdistance}.
In the new RG scale (dashed line), all we know is that the spins at
times \( \tau _{1} \) and \( \tau _{2} \) have the same orientation.
Each process compatible with the histories shown in the figure is
an independent part of the partition function. For definiteness, let
us assume that in this position there is a net charge \( \left( 0,1,1\right)  \).
At the new scale there are two indistinguishable possibilities to
be considered: either there is a single particle produced by a term
of \( H^{b}_{z} \), or there is a {}``close pair'' at \( \tau  \)
and \( \tau +\delta \tau  \) that was fused.

The effective Hamiltonian strategy is to reconstruct the CG at each
RG step. Since there is no spin flip between \( \tau _{1} \) and
\( \tau _{2} \) and there is a net charge \( \left( 0,1,1\right)  \),
the operator that performs this task is \begin{equation}
\label{ex1}
F_{L\uparrow }^{\dagger }F_{R\uparrow }^{\phantom {\dagger }}\bar{S}^{z}\left( \bar{x},\bar{\tau }\right) e^{i\sqrt{2\pi g_{\sigma }}\bar{\phi }_{s}\left( \bar{x},\bar{\tau }\right) +i\sqrt{2\pi g_{\rho }}\bar{\phi }_{c}\left( \bar{x},\bar{\tau }\right) },
\end{equation}
where the overbar denotes an operator at the new scale.

We want to know how to compare the spins at the new scale with the
ones at the previous scale. In the first history of Fig.~\ref{smallestdistance}
this is a trivial question. Before rescaling, the process had the
same form, so \[
\bar{S}^{z}\left( \bar{x},\bar{\tau }\right) =S^{z}\left( x,\tau \right) .\]

On the other hand, there are four possible bosonic operators that
can fit into the second history case . Let us start with the {}``close
pair''\begin{eqnarray*}
 &  & F_{L\uparrow }^{\dagger }F_{L\downarrow }^{\phantom {\dagger }}F_{L\downarrow }^{\dagger }F_{R\uparrow }^{\phantom {\dagger }}S^{+}\left( x,\tau +\delta \tau \right) S^{-}\left( x,\tau \right) \\
 &  & e^{i\sqrt{2\pi g_{\sigma }}\phi _{s}\left( x,\tau +\delta \tau \right) +i\sqrt{\frac{2\pi }{g_{\sigma }}\kappa }\theta _{s}\left( x,\tau +\delta \tau \right) }\\
 &  & e^{i\sqrt{2\pi g_{\rho }}\phi _{c}\left( x,\tau \right) -i\sqrt{\frac{2\pi }{g_{\sigma }}\kappa }\theta _{s}\left( x,\tau \right) }.
\end{eqnarray*}
Point-splitting the bosonic operators, we obtain

\[
F_{L\uparrow }^{\dagger }F_{R\uparrow }^{\phantom {\dagger }}S^{+}\left( x,\tau +\delta \tau \right) S^{-}\left( x,\tau \right) e^{i\sqrt{2\pi g_{\sigma }}\bar{\phi }_{s}\left( \bar{x},\bar{\tau }\right) +i\sqrt{2\pi g_{\rho }}\bar{\phi }_{c}\left( \bar{x},\bar{\tau }\right) }.\]
Another possible pair is\begin{eqnarray*}
 &  & F_{L\downarrow }^{\dagger }F_{R\uparrow }^{\phantom {\dagger }}F_{L\uparrow }^{\dagger }F_{L\downarrow }^{\phantom {\dagger }}S^{-}\left( x,\tau +\delta \tau \right) S^{+}\left( x,\tau \right) \\
 &  & e^{i\sqrt{2\pi g_{\rho }}\phi _{c}\left( x,\tau +\delta \tau \right) +i\sqrt{\frac{2\pi }{g_{\sigma }}\kappa }\theta _{s}\left( x,\tau +\delta \tau \right) }\\
 &  & e^{i\sqrt{2\pi g_{\sigma }}\phi _{s}\left( x,\tau \right) -i\sqrt{\frac{2\pi }{g_{\sigma }}\kappa }\theta _{s}\left( x,\tau \right) }.
\end{eqnarray*}
Reordering the Klein Factors and point-splitting again, we can rewrite
this pair as \begin{eqnarray*}
 & - & F_{L\uparrow }^{\dagger }F_{R\uparrow }^{\phantom {\dagger }}S^{-}\left( x,\tau +\delta \tau \right) S^{+}\left( x,\tau \right) \\
 &  & e^{i\sqrt{2\pi g_{\sigma }}\bar{\phi }_{s}\left( \bar{x},\bar{\tau }\right) +i\sqrt{2\pi g_{\rho }}\bar{\phi }_{c}\left( \bar{x},\bar{\tau }\right) }.
\end{eqnarray*}
The other two possibilities give the same contributions as these ones.
If we now identify\begin{eqnarray*}
2\bar{S}^{z}\left( \bar{x},\bar{\tau }\right)  & \equiv  & S^{+}\left( x,\tau +\delta \tau \right) S^{-}\left( x,\tau \right) \\
 & - & S^{-}\left( x,\tau +\delta \tau \right) S^{+}\left( x,\tau \right) ,
\end{eqnarray*}
we reconstruct Eq.~\ref{ex1}. Note the importance of the Klein factors
for this identification to hold.

\end{document}